\documentclass[11pt,fullpage]{article}

\usepackage{times,amsmath,epsfig}
\usepackage{graphicx}
\usepackage{placeins}
\usepackage{url}
\usepackage{subfig}
\usepackage{paralist}
\usepackage{caption}
\usepackage{multirow}
\usepackage{fullpage}

\newcommand{\eat}[1]{}

\newtheorem{ex}{Example}

\begin{document}
\date{February 2013}
% ****************** TITLE ****************************************

\title{A Survey on Array Storage, Query Languages, and Systems}

% ****************** AUTHORS **************************************


%\numberofauthors{2} %  in this sample file, there are a *total*

\author{
% 1st. author
Florin Rusu \hspace*{2cm} Yu Cheng\\
       \small{University of California, Merced}\\
       \small{5200 N Lake Road}\\
       \small{Merced, CA 95343}\\
       \small\texttt{\{frusu,ycheng4\}@ucmerced.edu}
}
\maketitle

%%%%%%%%%%%%%%%%%%%%%%%%%%%%%%%%%%%%%%%%%%%%%%%%%%%%%%%%%%%%%%%%%%%%%%%%%%%%%%%%%%
%\input{abstract}
\begin{abstract}

Since scientific investigation is one of the most important providers of massive amounts of ordered data, there is a renewed interest in array data processing in the context of Big Data. To the best of our knowledge, a unified resource that summarizes and analyzes array processing research over its long existence is currently missing. In this survey, we provide a guide for past, present, and future research in array processing. The survey is organized along three main topics. Array storage discusses all the aspects related to array partitioning into chunks. The identification of a reduced set of array operators to form the foundation for an array query language is analyzed across multiple such proposals. Lastly, we survey real systems for array processing. The result is a thorough survey on array data storage and processing that should be consulted by anyone interested in this research topic, independent of experience level. The survey is not complete though. We greatly appreciate pointers towards any work we might have forgotten to mention.

\end{abstract}

%%%%%%%%%%%%%%%%%%%%%%%%%%%%%%%%%%%%%%%%%%%%%%%%%%%%%%%%%%%%%%%%%%%%%%%%%%%%%%%%%%
%\input{introduction}
\section{Introduction}\label{sec:intro}

\textit{Big Data}~\cite{mckinsey} is the new buzz word in computer science as of 2012. There are new conferences organized specifically to tackle Big Data issues. Many classical research areas -- beyond data management and databases -- allocate significant attention to Big Data problems. And, most importantly, state governments provide unprecedented amounts of funds to support Big Data research~\cite{white-house}---at the end of the day, Big Data analytics played a significant role in the 2012 US presidential elections.

Scientific investigation represents one of the most important sources of Big Data. Science generate massive amounts of data through high-rate measurements of physical conditions, environmental and astronomical observations, and high-precision simulations of physical phenomena. These are made possible by the technological advancements in data acquisition equipment and the magnified density acquired by the modern storage devices which practically resulted in infinite storage capacity. Since scientific data are intrinsically ordered -- positional and/or temporal -- indexed arrays gained a lot of attention as a more appropriate data structure to represent scientific data---the ubiquitous unordered relational model made popular by business database systems cannot handle massive ordered data optimally.

From an inexperienced eye, array data processing might seem as a new research direction. At a closer look though, we found out that the interest in storing and managing ordered datasets is at least two decades old, dating back to early 1990's~\cite{maier:order}. Given the technological changes suffered by the computer industry over this time span, it is interesting to analyze how research ideas dating back to the early days of array processing survived and evolved to the current solutions.
\begin{itemize}
\item What did get implemented in real systems?
\item What ideas succumbed in complete oblivion?
\item What ideas survived and how did they change over time?
\item What ideas did get rediscovered?
\end{itemize}
These are all important questions that have to be addressed in order to understand the research problems specific to array data processing.

In this paper, we study the fundamental research directions in array data storage, query languages, and systems in the context of massive scale scientific data processing. Our definitive goal is to identify and analyze the most important research ideas proposed for each of these topics over time. The resulting survey is designed to serve two principal objectives. First, it summarizes accurately the most important ideas in array storage and processing by identifying the main research problems. And second, it organizes this material to provide an accurate perspective for the current interest in array processing. To the best of our knowledge, this is the first complete survey on array data storage and processing.

To this end, we survey a large body of work spread over more than two decades to find answers to the questions enumerated above. We start with a theoretical formalization of array data in Section~\ref{sec:arrays}. Array chunking techniques, chunk storage across a single and multiple disks, and chunk organization are presented to the deepest detail in Section~\ref{sec:array:storage}. Array algebras and array query languages are introduced in Section~\ref{sec:query-lang}. None of them has sufficient traction at this time. In Section~\ref{sec:array-syst}, we analyze the proposed systems for large scale array processing---past to present. We mostly focus on the execution strategies for primary array operations. We allocate ample space discussing the capabilities of the state-of-the-art array processing system -- SciDB~\cite{scidb:demo} -- in Section~\ref{sec:state-of-the-art}. The most recent research problems in the context of scientific data processing and how they translate to ordered datasets are also presented in Section~\ref{sec:state-of-the-art}. We conclude the survey by pointing out current research directions in Section~\ref{sec:conclusions}.

%%%%%%%%%%%%%%%%%%%%%%%%%%%%%%%%%%%%%%%%%%%%%%%%%%%%%%%%%%%%%%%%%%%%%%%%%%%%%%%%%%
%\input{arrays}
\section{Arrays}\label{sec:arrays}

Consider a multi-set of discrete domains $D_{i} = [l_{i}, u_{i}]$, $i \in \left\{1, 2, \dots, N\right\}$ where each domain $D_{i}$ contains integers between $l_{i}$ and $u_{i}$. An $N$-dimensional array with $M$ attributes $A_{j}$, $j \in \left\{1, 2, \dots, M\right\}$, can be thought of as a function defined over dimensions and taking values attribute tuples, i.e.,:
\begin{equation}\label{eq:array-def}
	Array : D_{1} \times D_{2} \times \dots \times D_{N} \longmapsto \left(A_{1}, A_{2}, \dots, A_{M}\right)
\end{equation}
where the type of the attributes can be any simple data type encountered in the relational data model. Using the same ideas behind extended data types, or user-defined data types, it is possible to have attributes with composite types, e.g., array, case in which we have nested arrays.

\subsection{Arrays and Relations}\label{ssec:array-vs-rel}

To better understand the difference between arrays and relations, it is important to clarify the distinction between dimensions and attributes. A relation can be viewed as an array without dimensions, only with attributes. Thus, there is no ordering function that allows the identification of a tuple based on dimensional indexes. Going from relations to arrays, it is required that dimension attributes form a key in the corresponding relation, i.e., there is a functional dependence from the dimension attributes to all the other attributes in the relation. Since a key is maximal, any attribute can be immediately transformed into a dimension. Transforming dimensions into attributes is not that straightforward. To be precise, converting a dimension into an attribute is equivalent to destroying the array property and losing any ordering information. As such, any array can be viewed as a particular type of relation organized along dimensions.

Essentially, the expression $\text{\textit{Array}}\left[d_{1}, d_{2}, \dots, d_{N}\right]$, where $d_{i} \in \left[l_{i}, u_{i}\right]$, makes sense for an array and is uniquely determined. The same is true for a relation in which $\left(d_{1}, d_{2}, \dots, d_{N}\right)$ represents a key. What distinguishes an array from a relation though is that the array is organized such that finding the entry $\text{\textit{Array}}\left[d_{1}, d_{2}, \dots, d_{N}\right]$ can be done directly from the value of the indexes (the position), without looking at any other entries. This is not possible in a relation since there is no correspondence between the indexes and the actual position in the physical representation---at least in the abstract relational model. Consequently, we find the main difference between arrays and relations at the physical organization level since arrays are a particular type of relation from an abstract perspective.

\subsection{Reducing the Dimensionality of an Array}\label{ssec:array:dim-red}

We have seen that converting a dimension into an attribute has the effect of destroying the array property, i.e., the array is up-casted to a relation. A different approach to reduce the dimensionality of an array is to simply eliminate a dimension. The immediate effect is that the remaining dimensions do not form a key anymore since as many duplicates as the size of the domain of the eliminated dimension are introduced. In order to preserve the array property even when dimensions are eliminated, the array has to \textit{split} or \textit{sliced} into multiple arrays with lower dimensionality. The number of these arrays is given by the size of the domain of the eliminated dimension. For example, let us assume that we reduce the dimensionality of $\text{\textit{Array}}$ by eliminating the first dimension $D_{1}$. The result is $u_{1} - l_{1} + 1$ arrays of dimensionality $N - 1$ given by:
\begin{equation}\label{eq:array:dim-red}
\begin{aligned}
	& Array_{l_{1}} : D_{2} \times \dots \times D_{N} \longmapsto \left(A_{1}, A_{2}, \dots, A_{M}\right)\\
	& Array_{l_{1} + 1} : D_{2} \times \dots \times D_{N} \longmapsto \left(A_{1}, A_{2}, \dots, A_{M}\right)\\
	& \dots \\
	& Array_{u_{1}} : D_{2} \times \dots \times D_{N} \longmapsto \left(A_{1}, A_{2}, \dots, A_{M}\right)
\end{aligned}
\end{equation}
Evaluating the expression $\text{\textit{Array}}\left[d_{1}, d_{2}, \dots, d_{N}\right]$ requires two steps in the new representation. First, the $d_{1}$ array has to be identified. Then, the expression $\text{\textit{Array}}_{d_{1}}\left[d_{2}, \dots, d_{N}\right]$ has to be evaluated instead. Dimensionality reduction can be generalized to any number of dimensions. The result is a large number of lower dimensionality arrays. Although the benefits of such a decomposition might not be visible immediately, there are classes of queries that benefit from this representation.

\subsection{Array Types}\label{ssec:array:types}

There are two types of array data -- dense and sparse -- classified according to the number of entries defined for the \textit{Array} function. If \textit{Array} is defined for each entry in the input domain, i.e., for each of the $|D_{1}| * |D_{2}| * \dots * |D_{N}|$ entries, then the array is considered \textit{dense}, also known as \textit{grid} or multidimensional discrete data (MDD)~\cite{furtado:tiling}. Grids contain values in each cell. As an example, consider a digitized 2-D image where the pixel at each position consists of three byte values, one corresponding to each R, G, B intensity, respectively.

\textit{Sparse arrays} can be thought of as incomplete grids with missing cells. Intuitively, sparse arrays are obtained by making the size of the domain for each dimension extremely large while providing values only for a limited number of cells. Consider, for example, the case of dimensions defined over real-valued domains. Notice that it is also possible to go the other direction---transform sparse arrays into grids. The idea is to condense many index values across each dimension into a single scalar such that all the cells contain at least one value. As a result though, it is very likely that a cell contains more than a single tuple, case in which there are two alternatives---store the tuples with all or a part of their attributes independently or create a single tuple that aggregates multiple values in each cell. This process corresponds to histogram binning in approximate query processing.

The intuitive way to understand the difference between dense and sparse arrays is to look at the expression $\text{\textit{Array}}\left[d_{1}, d_{2}, \dots, d_{N}\right]$. In the case of a grid, this expression always returns a valid cell containing data. That is not the case for sparse arrays. It is very likely that in a sparse array the cell is empty and does not contain valid data. As a result, the strategies to store the two types of arrays are quite different.

%%%%%%%%%%%%%%%%%%%%%%%%%%%%%%%%%%%%%%%%%%%%%%%%%%%%%%%%%%%%%%%%%%%%%%%%%%%%%%%%%%
%\input{array-storage}
\section{Array Storage}\label{sec:array:storage}

Array storage has to be considered in the following context. The size of the array, $|D_{1}| * |D_{2}| * \dots * |D_{N}| * \left|\text{\textit{sizeof}}\left(A_{1}, A_{2}, \dots, A_{M}\right)\right|$, is too large to fit entirely in memory. Nonetheless, it is possible to access the array elements based on their index. Array elements are organized on disk into fixed size \textit{blocks} or \textit{chunks} that contain a group of array elements. Let us consider the size of a chunk to be $B$ bytes. Whenever an element from a chunk has to be read into memory, the entire chunk is read---the I/O unit is the chunk, similar to the page for file systems and the block for relational databases, respectively. Based on these considerations, the entire array storage problem reduces to a series of questions:
\begin{enumerate}
\item What value is $B$ taking?
\item How to decide what array cells are put together in the same chunk? Or, equivalently, what is the shape of the chunk?
\item What is the mapping function from an array index to the corresponding chunk on disk?
\item How to organize the cells inside the chunk?
\end{enumerate}
Different chunking strategies answer these questions in different ways. We consider the most important strategies in details in the following. We start with general strategies that can be applied equally to dense and sparse arrays. Then we consider more specific strategies.

\subsection{Chunk Size $B$}\label{ssec:array:storage:chunk-size}

Before that though, we can answer the first question in the general case. In early work~\cite{sarawagi:chunking}, it was common to set $B$ to the size of the file system page or the database block size, e.g., 4 to 64 KB. This strategy keeps the chunks tight without wasting space due to fragmentation. It is also optimal when small portions of the array are retrieved by the majority of the accesses, e.g., direct cell access based on the indexes or selective range queries. In more recent work~\cite{arraystore}, $B$ is set to much larger values, in the order of megabytes or even tens of megabytes. There are two main reasons for this. First, scanning larger continuous portions from the disk up to these sizes does not take considerably longer (due to the logic implemented in the disk controller). Repositioning the disk reading head for small requests is still the bottleneck. And second, memory capacity increased considerably, thus allowing for more data to be stored in memory. And if the memory segments are continuous it is even better.

\subsection{Arbitrary Chunking}\label{ssec:array:storage:arbitrary}

Arbitrary chunking is the most straightforward chunking strategy. It does not require any mathematical formulation or any other kind of information. The main idea behind this strategy is to group together in the same chunk cells that are close to each other. Closeness is measured based on dimensions. A common simplification that is applied is to enforce that the shape of the resulting chunks is a multi-dimensional hyper-cube aligned with the dimension axes. Then, the questions that have to be answered are where to position the hyper-planes corresponding to each axis? And are these hyper-planes bounded or unbounded, i.e., do they cover the entire axis or only a segment?

\textit{Regular chunking}~\cite{arraystore} or \textit{aligned tiling}~\cite{furtado:tiling} provides the simplest answers to these two questions. Each dimension is divided into equal segments. And the segments cover the entire axis. The result is a set of identical hyper-cubes aligned with the axes. A chunk corresponds to each such hyper-cube. In regular chunking, the number of hyper-planes on each axis is chosen arbitrarily. In aligned tiling, it is chosen such that the resulting chunks represent a uniform scaling down of the entire domain that fits in the allocated chunk size $B$, i.e., the ratio between the chunk size and the domain size is identical on all dimensions.

\begin{ex}[Regular chunking]\label{ex:reg-chunking}
Consider a 3-D grid of integers. The domain sizes along the 3 dimensions are $\left(7,500, 7,500, 20\right)$. Aligned tiling of this dense array requires chunk sizes that have the same ratio across all the dimensions. Thus, if we consider the constant ratio to be 10, then the chunk shape is $\left(750, 750, 2\right)$ and we get 1,000 chunks. Regular chunking does not require the same ratio. For example, chunks with the shape $\left(750, 375, 4\right)$ have different ratios on each dimension. Notice though, that the ratio is still an integer.
\end{ex}

In \textit{directional tiling}~\cite{furtado:tiling}, each dimension is treated independently. The position of the hyper-planes is given for each dimension. Chunks are obtained at the intersection of the hyper-planes. They will not necessarily contain the same number of points, i.e., they do not necessarily have the same shape. Chunks are aligned and \textit{irregular}. Careful consideration is required for the cases when the volume of a chunk is smaller than the maximum allowed volume $B$ -- merging is possible -- and when the volume is greater---further splitting is definitely required. When any of these operations is applied, chunks become \textit{nonaligned}, i.e., the hyper-plane is only a segment that does not cover the entire axis domain.

A special case of arbitrary chunking corresponds to \textit{slicing} a particular dimension with hyper-planes at every position in its domain. The resulting hyper-cubes have dimensionality $N$-1 and they can be chunked further independently of each other. Any processing that can be confined to a slice becomes simpler due to the reduced dimensionality. Processing across multiple slices has to be decomposed into separate processing on each slice---a loop over the slices. The default representation of multi-dimensional arrays in general-purpose programming languages, e.g., C, Java, is based on slicing. Starting with the most significant dimension, arrays of lower dimensionality are obtained by fixing the value of the outer indexes. Due to the linear representation in memory, these arrays are straightforward to generate. Problems appear when a lower-dimensional array has to be obtained by fixing the value of an index that does not match the linearization order. In this case, the lower-dimensional array has to be explicitly created by individually accessing each element. Consider, for example, a C language 2-D matrix \texttt{matrix[10][10]} linearized in standard row-major. Accessing the $3^{rd}$ row is straightforward, i.e., \texttt{matrix[2]}, but accessing the $3^{rd}$ column is impossible without explicitly accessing each element.

\begin{ex}[Sliced chunking]\label{ex:slice-chunking}
If we consider the same setting as in Example~\ref{ex:reg-chunking}, we can generate sliced chunks by treating each of the 20 points on the $3^{rd}$ dimension separately. We first generate 20 2-D arrays with size $\left(7,500, 7,500\right)$. Then, we can chunk each of them individually.
\end{ex}

\subsection{Workload-based Chunking}\label{ssec:array:storage:workload}

The actual storage organization of an array is strongly dependent on the access patterns used to access the cells. It is always the case that the access patterns are application and workload specific. Thus, there is no organization that provides optimal performance for all the possible queries. In the worst case, the entire array has to be read from disk in order to compute the result. This is equivalent to a complete table scan in the relational model. In the best case, only those chunks containing data relevant to the query at hand are read from disk. Given these two extremes, the actual storage organization has to minimize the number of blocks read from disk for the majority of queries in the workload. A second parameter that requires careful consideration when deciding upon the partitioning strategy is the actual size of the query result. It is very likely that in the case of queries returning a large number of cells the difference between strategies is not that significant. When only a handful of cells are returned though, the actual partitioning strategy plays a very important role. This problem is closely related to the effectiveness of indexes in relational databases.

In~\cite{furtado:tiling}, the authors identify a set of frequent access patterns specific for multi-dimensional arrays. We summarize them in the following:
\begin{itemize}
\item \textit{Subsample multi-dimensional area with same dimensionality.} The result of such a query is a hypercube having the same dimensionality as the original array. Splitting the array into blocks across all the dimensions is the optimal strategy in this situation. Notice that accessing the whole array is a particular case of this access pattern.
\item \textit{Section of lower dimensionality across a subset of dimensions}. In this case, the query result is typically a hyperplane with a lower dimensionality. The storage organization matching the section pattern provides optimal access in this case.
\end{itemize}
The authors propose a chunking strategy that takes the workload queries into account. Based on the query log, the access frequency is measured for each hyper-cube of the array. Hyper-cubes accessed frequently enough are designated \textit{areas of interest}. Directional tiling is directly applied by taking the sides of the areas of interest as the splitting hyper-planes across each dimension. Further partitioning or merging might be required, as in directional tiling. Merging is different due to the requirement that only tiles from the same area(s) of interest can be put together. The objective is to put as much data as possible together from a single area of interest and to minimize the number of tiles for a given area of interest.

In~\cite{arraystore}, the authors provide more varied access patterns in addition to the range selection patterns across all or a subset of dimensions introduced previously. These more varied access patterns are:
\begin{itemize}
\item \textit{Structural join between two arrays.} This operation requires combining data from the same index position in two arrays having the same dimensionality. The straightforward (and optimal) solution is to partition the arrays identically and, in the case of parallel processing, store corresponding partitions on the same processing node.
\item \textit{Overlap operations that access adjacent cells.} If the adjacent cells that are accessed are confined to a bounded region, the data can be duplicated at each array partition. This allows for independent parallel processing at each data partition without communication across different partitions. If the number of adjacent cells is not bounded, merging and communication across array partitions are required---this is the more general solution.
\end{itemize}
The chunking strategies proposed by the authors to handle these two access patterns are variations on regular and directional tiling. The main idea is to apply \textit{two-level regular or directional tiling}. At the upper level, the chunk is determined as in any of these strategies. Inside a chunk -- the lower level -- another chunking is executed using again one of the regular or directional algorithms. The mini-chunks resulted at the lower level can be accessed as a unit of processing---the I/O unit is still the upper level chunk. This strategy is beneficial exactly when data from adjacent chunks is needed. Instead of passing the entire chunk, only the overlapping mini-chunks have to be transferred to an adjacent chunk. In our opinion, the improvement generated by this two-level strategy is questionable given that the entire chunk still needs to be read from disk---and this is the dominating cost, not the processing. The second chunking strategy proposed for overlapped processing requires duplicating data across multiple chunks. This \textit{materialized view} can be stored either with the main chunk or separated. Additional space is used in both cases. The advantage of storing the materialized view separated is that it can be read on demand, only when needed. Otherwise, it has to be read whenever the chunk is read and this has the potential to incur significant overhead. As a variation on the same idea, multiple concentric materialized views with increasing radius can be created. The exterior ones always include the interior ones.

It is important to remark that these access patterns are based entirely on dimensions. Whenever the array has to be accessed based on attribute values, inspecting all the cells represents the only alternative. This corresponds to reading all the data in the worst case. If columnar storage is used in array partitions though, the amount of data read from disk is drastically reduced---only the needed attributes are read. Another alternative to reduce the amount of data read from disk is to build an unclustered index which stores for each distinct value of the attribute the block(s) where it appears. Nonetheless, this solution requires additional space for storing the index and some additional time to access the index prior to access the data (as with any index structure).

\subsection{Workload-based Chunking using Optimization Formulations}\label{ssec:array:storage:workload:optim}

\paragraph{Query shape model.}
The approach taken in~\cite{sarawagi:chunking} is to model all of the observable access patterns as a probability distribution function over the shapes of the accesses---query shape model~\cite{rotem:optimal-chunking}. Essentially, accesses are represented as N-dimensional hyper-cubes with a corresponding length in each dimension, i.e., $\left(s_{1}, s_{2}, \dots, s_{N}\right)$. While this is the most general form of access, notice that it also encompasses degenerated patterns such as accessing a single cell or a hyper-plane, i.e., $s_{i} = 1$. A probability is assigned to each access pattern independent of the actual occurring position in the array---the positions are assumed to be uniformly distributed across the entire domain. Access patterns can then be grouped into classes of the form $\left\{\left[P_{i}, \left(s_{i_{1}}, s_{i_{2}}, \dots, s_{i_{N}}\right)\right] : 1 \leq i \leq K\right\}$ where $K$ is the number of different hyper-cube shapes, i.e., classes, and $P_{i}$ is the probability corresponding to each class. In order to determine the optimal chunking -- only regular chunking is considered -- an optimization problem to minimize the number of blocks read from disk across all the classes is formulated:

\textit{Sarawagi-Stonebraker (SS) formulation~\cite{sarawagi:chunking}}
\begin{equation}\label{eq:sarawagi:optim-form}
\begin{aligned}
	& \text{min}_{\left(c_{1}, c_{2}, \dots, c_{N}\right)} \sum_{i=1}^{K}\left(\prod_{j=1}^{N}\left\lceil\frac{s_{i_{j}}}{c_{j}}\right\rceil\right) P_{i}\\
	& \text{subject to: } \prod_{i=1}^{N} c_{i} \leq \frac{B}{\left|\text{\textit{sizeof}}\left(A_{1}, A_{2}, \dots, A_{M}\right)\right|}\\
	& \text{where } \left(c_{1}, c_{2}, \dots, c_{N}\right) \text{ is the shape of the regular chunk}
\end{aligned}
\end{equation}
The only constraint is given by the size of the disk block and the requirement that a chunk has to fit in a single block. The authors make the assumption that, independent of the relative position of the chunk and the query hyper-cube, at most one additional chunk is read from disk for each dimension---that is where the ceiling comes from in the objective function. Clearly, this assumption is dependent on the actual position of the chunk (the shape of the chunk) and what range query has to be answered---in some cases, two additional chunks have to be read. What amplifies the error effect is the multiplication of the factors across dimensions while assuming dimensionality independence. Thus, the error becomes significantly higher if the same error is made for all the dimensions of a given class---the higher the dimensionality of the array, the higher the error. As a result, the solution of this formulation is only approximate and can incur significant errors.

In~\cite{rotem:optimal-chunking}, the authors modify the objective function of the optimization formulation exactly by observing that the assumption made in~\cite{sarawagi:chunking} on the number of chunks to be read is problematic. The changed objective function is given below:

\textit{Expectation formulation~\cite{rotem:optimal-chunking}}
\begin{equation}\label{eq:rotem:optim-form}
\begin{aligned}
	& \text{min}_{\left(c_{1}, c_{2}, \dots, c_{N}\right)} \sum_{i=1}^{K}\left(\prod_{j=1}^{N}\left(\frac{s_{i_{j}} - 1}{c_{j}} + 1\right)\right) \end{aligned}
\end{equation}
The factors in this formulation represent the expected value of the number of chunks to be read for each query class under the assumption that the position of the queries is uniformly distributed over the entire array domain. Notice that the objective function takes real values in this case. These values do not represent the exact number of chunks to be read from disk for a given set of queries.

No matter which formulation we consider, it is not possible to compute a closed-form solution. Since it is not feasible to search the entire solution space, i.e., $|D_{1}| * |D_{2}| * \dots * |D_{N}|$, methods to prune the space are required. The solution proposed in~\cite{sarawagi:chunking} reduces the solution space to shapes of the form $\left(2^{y_{1}}, 2^{y_{2}}, \dots, 2^{y_{N}}\right)$ that are maximal, i.e., $\sum_{i=1}^{N}y_{i} = \left\lceil\log_{2}\frac{B}{\left|\text{\textit{sizeof}}\left(A_{1}, A_{2}, \dots, A_{M}\right)\right|}\right\rceil$, where $y_{i}$, $1 \leq i \leq N$, are positive integers. Essentially, only hyper-cubes with side lengths powers of 2 are considered that fill the maximum chunk size with minimal waste. The search over such shapes is exhaustive. Once the optimal shape with this restricted form is found, an additional search around the solution can be triggered to find an even better solution that allows more general shapes. In~\cite{rotem:optimal-chunking}, the exhaustive search at the first level is replaced with a Greedy algorithm that starts with 0 lengths for all dimensions and then chooses optimally which dimension to increase at each step. The computations required at each step are intricate and the authors do not show what benefits this brings when compared to exhaustive search over the parameter space. While the solution to the optimization problem is guaranteed to provide the optimal chunking under the chosen assumptions, it is interesting to see how far away is from the best solution. None of~\cite{sarawagi:chunking,rotem:optimal-chunking} provide such results or mention if other chunking strategies are better for the query classes used in experiments.

\paragraph{Independent attribute range model.}
The query workload can also be modeled through the size of the ranges it contains on each dimension. Instead of considering a hyper-cube as the access unit, we decompose the query in its corresponding segments on each dimension. Thus, two 2-D query shapes $\{<4, 4>, ,4, 6>\}$ are part of two different pattern classes in the query shape model, but they are part of the same class based on the first dimension $D_{1}$. In this case, the probability distribution is defined separately for each dimension. If the distributions are assumed independent across the dimensions, the model is called independent attribute range model~\cite{rotem:optimal-chunking}. We can formulate an optimization problem that minimizes the number of chunks read from disk given a query workload modeled using the independent attribute range model in a similar fashion to the query shape model:

\textit{Independent attribute range formulation~\cite{rotem:optimal-chunking}}
\begin{equation}\label{eq:iar:optim-form}
\begin{aligned}
	& \text{min}_{\left(c_{1}, c_{2}, \dots, c_{N}\right)} \prod_{i=1}^{N}\left[\sum_{j=1}^{m_{i}}\left(\frac{s_{i_{j}} - 1}{c_{i}} + 1\right) P_{i_{j}}\right]\\
	& \text{subject to: } \prod_{i=1}^{N} c_{i} \leq \frac{B}{\left|\text{\textit{sizeof}}\left(A_{1}, A_{2}, \dots, A_{M}\right)\right|}\\
	& \text{where } \left(c_{1}, c_{2}, \dots, c_{N}\right) \text{ is the shape of the regular chunk},\\
	& \hspace*{1cm} m_{1}, m_{2}, \dots, m_{N} \text{ are the number of range sizes on each dimension},\\
	& \hspace*{1cm} s_{i_{j}}, \text{ is the } j^{\text{\textit{th}}} \text{ range on dimension } i,\\
	& \hspace*{1cm} P_{i_{j}}, \text{ is the probability of the } j^{\text{\textit{th}}} \text{ range on dimension } i, \sum_{j=1}^{m_{i}}P_{i_{j}} = 1, 1 \leq i \leq N
\end{aligned}
\end{equation}
A closed form formula can be computed if we give up the requirement that $c_{i}$'s have to be positive integers and we impose maximality in the sense defined for the query shape model. The formula obtained using the Lagrange multiplier method gives us the $c_{i}$'s:
\begin{equation}\label{eq:iar:optim-form:solution}
\begin{aligned}
	& c_{i} = \bar{A_{i}} \left(\frac{\frac{B}{\left|\text{\textit{sizeof}}\left(A_{1}, A_{2}, \dots, A_{M}\right)\right|}}{\prod_{i=1}^{N}\bar{A_{i}}}\right)^{\frac{1}{n}}, 1 \leq i \leq N\\
	& \text{where } \bar{A_{i}} = \sum_{j=1}^{m_{i}} s_{i_{j}} P_{i_{j}} - 1
\end{aligned}
\end{equation}
In order to obtain the integral solution, i.e., positive integer values for $c_{i}$'s, the authors propose a method that rounds up some of the $c_{i}$'s while the others are rounded down. By carefully choosing the two partitions, the integral optimal solution can be obtained.

\subsection{Mapping Array Cells to Chunks}\label{ssec:array:storage:mapping-cells}

Independent of the actual chunking strategy, neighboring cells in the case of dense grids or close points in sparse arrays -- measured based on dimensions -- are grouped together in chunks. The shape of the chunk is always a hyper-cube aligned with the dimensions. Whenever a cell or a range of cells have to be retrieved based on their index, the chunk(s) containing the cell(s) have to be found first. This requires a mapping function from the cell index to the corresponding chunk:
\begin{equation}\label{eq:implicit-mapping}
\begin{aligned}
	& \text{\textit{Mapping}}_{\text{\textit{implicit}}} : D_{1} \times D_{2} \times \dots \times D_{N} \longmapsto \left[1 \dots Z_{\text{\textit{chunks}}}\right],\\
	& \text{where } Z_{\text{\textit{chunks}}} \text{ is the maximum number of chunks}
\end{aligned}
\end{equation}
A chunk is identified by its position across all the chunks---chunks are linearized similarly to how they are stored on disk. The position is given as a number on a discrete axis. The mapping function can be implicit, given by a closed-form formula, or it can be explicit, stating for each chunk hyper-cube its corresponding position. An \textit{implicit mapping} requires a pre-determined order in which chunks are considered in the dimension space, e.g., row-major or column-major for 2-D arrays. The order impacts how chunks are retrieved from disk. To be precise, the order determines how long are the sequential scans. An implicit mapping function requires regular chunks having the same size and can be applied only to dense grids. When applied to sparse arrays, all the empty cells have to be represented explicitly, i.e., NULLs, thus wasting significant storage space.

\begin{ex}[Implicit mapping]\label{ex:implicit-mapping}
Consider the same 3-D grid as in Example~\ref{ex:reg-chunking}, i.e., $\left(7,500, 7,500, 20\right)$, chunked into 1,000 regular tiles with shape $\left(750, 750, 2\right)$. The mapping function used to linearize the chunks on disk considers the dimensions from the first to the third, with the index on the third increasing the fastest, i.e., row-major extended to three dimensions. Thus, the order in which chunks are stored on disk follows the indexes as: $\left(1, 1, 1\right), \left(1, 1, 2\right), \dots, \left(1, 1, 10\right), \left(1, 2, 1\right), \dots, \left(1, 10, 10\right), \left(2, 1, 1\right), \dots$. It is straightforward to determine the formula for the mapping function in this case as:
\begin{equation}\label{eq:implicit-mapping:ex}
\begin{aligned}
	& \text{\textit{Mapping}}_{\text{\textit{implicit}}} (x, y, z) = \left\lfloor \frac{x}{750} \right\rfloor * 10 * 10 + \left\lfloor \frac{y}{750} \right\rfloor * 10 + \left\lceil \frac{z}{2} \right\rceil
\end{aligned}
\end{equation}
Given an array index, it is straightforward to find the chunk that contains the corresponding cell. For example, the cell corresponding to index $\left(1111, 308, 7\right)$ is in chunk $\left\lfloor \frac{1111}{750} \right\rfloor * 10 * 10 + \left\lfloor \frac{308}{750} \right\rfloor * 10 + \left\lceil \frac{7}{2} \right\rceil = 100 + 0 + 4 = 104$. Identifying the chunks corresponding to a range query is a bit more intricate. Consider the range $\left(\left[3,000, 4,000\right], \left[1,000, 7,000\right], \left[5, 11\right]\right)$. We have to treat each dimension separately in order to identify the range of chunks covered by the query interval. For dimension $x$, we have $\left[4, 5\right]$, for $y$, $\left[1, 9\right]$, and $\left[3, 6\right]$ on z, respectively. To find the chunks that overlap this range, we have to compute all the possible combinations of indexes, i.e., $2 * 9 * 4 = 72$. Thus, there are 72 chunks that have to be read in order to answer this range query. Some of them are: 413, 414, 415, 416, $\dots$, 445, 446, $\dots$, 593, 594, 595, 596.

It is interesting to evaluate the effectiveness of the chosen chunking scheme. For this, we compute the number of chunks that contain all the cells retrieved by the range query---we assume the same shape of the chunk. There are $1,001 * 6,001 * 7 = 42,049,007$ cells covered by the query. They fit in 38 chunks. With regular chunking, we have to read 72. That is almost double!
\end{ex}

An alternative approach is to first partition the array into regular chunks using any of the methods presented previously. The domain on each dimension is reduced from the original size to the number of chunks along the dimension. Based on the original dimension ordering, a chunk corresponds to each point in the new domain. A chunk can be identified by its position along each dimension. Given the new coordinate system, the mapping function is defined as a dimensionality reduction transformation from the chunk multi-dimensional coordinates to an integral position along a linear axis. Essentially, we are introducing an intermediate mapping from the original domain to the chunk domain, in the same multi-dimensional space. Only then we are mapping the chunks to the linear axis. Formally, this corresponds to two functions:
\begin{equation}\label{eq:mapping-chunk-linear}
\begin{aligned}
	& \text{\textit{Mapping}}_{\text{\textit{chunk}}} : D_{1} \times D_{2} \times \dots \times D_{N} \longmapsto C_{1} \times C_{2} \times \dots \times C_{N}\\
	& \text{\textit{Mapping}}_{\text{\textit{linear}}} : C_{1} \times C_{2} \times \dots \times C_{N} \longmapsto \left[1 \dots Z_{\text{\textit{chunks}}}\right],\\
	& \text{where } C_{i}, 1 \leq i \leq N, \text{ are the number of chunks along each dimension}, \\
	& \hspace*{1cm}Z_{\text{\textit{chunks}}} = |C_{1}| * |C_{2}| * \dots * |C_{N}| \text{ is the total number of chunks}
\end{aligned}
\end{equation}
While $\text{\textit{Mapping}}_{\text{\textit{chunk}}}$ is straightforward to define, there are a variety of choices for $\text{\textit{Mapping}}_{\text{\textit{linear}}}$. The most common choices are row-major (column-major), snake row-major (snake column-major)~\cite{jagadish:hilbert-clustering}, and their extensions to multi-dimensional spaces. Mapping functions based on space-filling curves are presented in~\cite{jagadish:hilbert-clustering}. They are defined recursively for a given domain size and do not have a simple closed-form formula. Out of the three methods presented -- Z-curve, Gray code, and Hilbert curve -- it is shown in~\cite{jagadish:hilbert-clustering} that Hilbert curve mapping provides the best performance for partial exact match selection -- slicing along one dimension -- and range selections in 2-D space. The performance metric used for the theoretical analysis is the number of runs of consecutive grid points which is equivalent to the number of non-consecutive disk blocks fetched. Lower values correspond to a reduced number of disk seek jumps. This translates indirectly to continuous scans, thus better disk I/O throughput. In addition to this metric, the total number of disk blocks fetched and the size of the linear span for a given selection -- the difference between the maximum and minimum linear coordinate -- are also used in the experimental evaluation.

\paragraph{Dimension ordering.}
An important question that requires attention in the case of implicit mapping is the order in which to consider the dimensions when linearizing the chunks on disk. Notice that the same number of chunks has to be read from disk no matter what the order is. What is highly sensitive to the order though is the length of the sequential scans and that of the seeks between chunks that are within query range. Longer sequential scans and shorter seek jumps are better. The arbitrary -- and most common -- solution is to use the order in which dimensions are specified in the array definition. In~\cite{sarawagi:chunking}, the authors provide an heuristic which orders the dimensions based on the ratio of the number of chunks read on a dimension -- across the queries in the workload -- and the number of chunks on that dimension. The dimension with the largest ratio is the inner-most one. Intuitively, this corresponds to having the dimensions with the largest number of chunks read in the inner loops of the traversing order. Or, equivalently, make the longer sequential scans more often and the longer seek jumps less frequent.

An \textit{explicit mapping} function bypasses the conversion to a chunk position and maps N-dimensional hyper-cubes specified by their left-bottom and right-upper corners, respectively, to the starting position of the corresponding chunk on disk:
\begin{equation}\label{eq:explicit-mapping}
\begin{aligned}
	& \text{\textit{Mapping}}_{\text{\textit{explicit}}} : \left[C_{l_{1}}, C_{u_{1}}\right] \times \left[C_{l_{2}}, C_{u_{2}}\right] \times \dots \times \left[C_{l_{N}}, C_{u_{N}}\right] \longmapsto \left[1 \dots Z_{\text{\textit{disk}}}\right],\\
	& \text{where } Z_{\text{\textit{disk}}} \text{ is the maximum size of the disk}
\end{aligned}
\end{equation}
Finding the array cell corresponding to a given index requires in this case identifying the chunk which contains the index. Since the mapping function is represented explicitly, this reduces to inspecting each entry and checking inclusion. Building a multi-dimensional index over the hyper-cubes is a possible alternative to reduce the number of entries inspected when the number of chunks is large. Irregular chunks require the mapping function to be represented explicitly.

\begin{ex}[Explicit mapping]\label{ex:explicit-mapping}
Let us consider a modification of the 3-D array used in the previous examples. Instead of having a dense grid, consider the $\left(7,500, 7,500\right)$ squares positioned in a plane of size $\left(10^{6}, 10^{6}\right)$. The resulting 20 2-D arrays, when considered together over the $\left(10^{6}, 10^{6}, 20\right)$ space, form a sparse array. One strategy to chunk the sparse array in this case is to slice each square out and chunk it using a dense strategy. We still want to represent the absolute indexes of each chunk though. Thus, if we use $\left(750, 750\right)$ rectangles as before, we obtain 2,000 chunks with 6 coordinates represented for each chunk. They have the form $\left[\left(x, y, z_{1}\right), \left(x + 750, y + 750, z_{1}\right)\right]$ or, equivalently, $\left[x, x + 750\right] \times \left[y, y + 750\right] \times \left[z_{1}, z_{1}\right]$. An explicit mapping function stores for each such hyper-cube the corresponding position on disk -- in a file that stores the entire array -- without first mapping to the chunk position. The reason for this sort of bypassing might not be evident in this example since chunks are regular.

The next question is how to handle this range information to optimally answer queries. For that, let us consider how point and range queries are answered. In both cases, the chunks that contain the point or overlap the range, respectively, have to be identified across all the chunks. The simple solution is to check each and every chunk---linear scan over all the chunks. A more intricate -- and likely more efficient -- solution is to build a spatial index such as R-tree across the hyper-cubes delimiting the chunks. Both point and range queries are handled efficiently by the index. The number of chunks that need to be inspected in this case is significantly reduced at the cost of building and maintaining the index.
\end{ex}

\subsection{Mapping Chunks to Disks}\label{ssec:array:storage:mapping-chunks}

A related problem that has to be considered in a multi-disk environment -- shared-disk or shared-nothing architecture -- is how to distribute chunks across disks or \textit{declustering}. The formal representation of this problem requires a function to be defined from the chunk linearization to the disk domain:
\begin{equation}\label{eq:disk-mapping}
\begin{aligned}
	& \text{\textit{Mapping}}_{\text{\textit{disk}}} : \left[1 \dots Z_{\text{\textit{chunks}}}\right] \longmapsto \left[1 \dots {\text{\textit{HD}}}\right], \\
	& \text{where } Z_{\text{\textit{chunks}}} \text{ is the maximum number of chunks}, \\
	& \hspace*{1cm}\text{\textit{HD}} \text{ is the maximum number of disks}
\end{aligned}
\end{equation}
Essentially, $\text{\textit{Mapping}}_{\text{\textit{disk}}}$ partitions the chunks over the available disks. The objective is to find such mappings that evenly distribute the chunks across all the available disks. This results in spreading the disk I/O evenly across disks, thus maximizing the overall throughput. While this can be achieved on the average, there will always be queries for which more chunks, if not all, have to be read from the same disk, resulting in degraded I/O performance. In order to get access to a given array cell, two mappings have to be applied in sequence, i.e., mapping composition. First, the chunk that contains the array cell has to be identified using either $\text{\textit{Mapping}}_{\text{\textit{implicit}}}$ or $\text{\textit{Mapping}}_{\text{\textit{explicit}}}$. Then, $\text{\textit{Mapping}}_{\text{\textit{disk}}}$ is applied on the result. The same procedure is followed for range queries, with individual calls to $\text{\textit{Mapping}}_{\text{\textit{disk}}}$ for each chunk in the overlapped region.

\paragraph{Data partitioning.}
In the following, we introduce possible forms for $\text{\textit{Mapping}}_{\text{\textit{disk}}}$ as presented in~\cite{arraystore,cart-prod-file:declustering}. It is important to remark that a significant number of the formulas are immediate extensions from data partitioning schemes in parallel databases. The main difference is that in data partitioning there is no mapping from an array cell to a given chunk, but rather the mapping is from a tuple attribute to a chunk, if present at all. The following mappings are inherited from parallel databases:
\begin{itemize}
\item \textit{Round-robin.} The mapping function is defined as $\text{\textit{Mapping}}_{\text{\textit{round-robin}}}\left(x\right) = \left(x + c\right) \mod \text{\textit{HD}} + 1$, where $c$ is a constant. The idea is to assign chunks sequentially to disks based on their position in the linear order. The distance between two chunks assigned to the same disk is $\text{\textit{HD}}$. Each disk receives at least $\frac{Z_{\text{\textit{chunks}}}}{\text{\textit{HD}}}$ chunks.
\item \textit{Range.} In range-based partitioning, the chunks are split into $\text{\textit{HD}}$ groups, each group containing $\frac{Z_{\text{\textit{chunks}}}}{\text{\textit{HD}}}$ chunks---we assume that $Z_{\text{\textit{chunks}}}$ is a multiple of $\text{\textit{HD}}$. This is similar to the round-robin scheme. The difference is that the groups contain consecutive chunks, i.e., $\text{\textit{Mapping}}_{\text{\textit{range}}}\left(x\right) = \left\lceil\frac{x * \text{\textit{HD}}}{Z_{\text{\textit{chunks}}}}\right\rceil$, where the division is integral.
\item \textit{Hash or pseudo-random.} The standard mapping function used in hash-based partitioning is given as $\text{\textit{Mapping}}_{\text{\textit{hash}}}\left(x\right) = \left[\left(a * x + b\right)\mod P\right]\mod \text{\textit{HD}} + 1$, where $a$ and $b$ are random numbers while $P$ is a large prime number. On average, the same number of chunks are assigned to each disk. What chunks get assigned to each disk though, depends strictly on the parameters. In order to enforce that the chunks are uniformly distributed across disks, a combination between round-robin and hash can be devised such that in each run of $\text{\textit{HD}}$ chunks, each disk gets one chunk. Inside a run, the assignment of chunks to disks is random rather than following a fixed pattern.
\end{itemize}

\paragraph{Declustering.}
Instead of assigning chunks to disks based on their linearization given either by the implicit or explicit mapping, the assignment can be done starting from the intermediate mapping $\text{\textit{Mapping}}_{\text{\textit{chunk}}}$. The input to $\text{\textit{Mapping}}_{\text{\textit{disk}}}$ is in this case the multi-dimensional coordinate in the chunk space:
\begin{equation}\label{eq:disk-mapping-from-chunk}
\begin{aligned}
	& \text{\textit{Mapping}}_{\text{\textit{disk-chunk}}} : C_{1} \times C_{2} \times \dots \times C_{N} \longmapsto \left[1 \dots {\text{\textit{HD}}}\right], \\
	& \text{where the symbols have the same meaning as previously defined}
\end{aligned}
\end{equation}
There are various forms $\text{\textit{Mapping}}_{\text{\textit{disk-chunk}}}$ can take. We introduce the most common declustering methods that use the intermediate mapping as presented in~\cite{cart-prod-file:declustering}:
\begin{itemize}
\item \textit{Disk modulo (DM).} In the DM scheme, chunk $\left[i_{1}, i_{2}, \dots, i_{N}\right]$ is assigned to disk $\left(i_{1} + i_{2} + \dots + i_{N}\right) \mod \text{\textit{HD}}$. Even though the assignment might seem simple, the DM mapping is known to be strictly optimal -- exactly the minimum number of chunks is read from each disk -- for many cases of partial match queries including all partial match queries with only one unspecified attribute~\cite{cart-prod-file:declustering}. Disk modulo does not scale as the number of disks is increased for range queries in particular. This limits drastically its applicability.
\item \textit{Fieldwise XOR (FX).} The FX scheme replaces the summation operation in DM with a bitwise XOR operation on the binary representation of the chunk coordinates. Chunk $\left[i_{1}, i_{2}, \dots, i_{N}\right]$ is assigned to disk $\left(i_{1} \oplus i_{2} \oplus \dots \oplus i_{N}\right) \mod \text{\textit{HD}}$, where $i_{j}$, $1 \leq j \leq N$, are binary representations of the indexes in the chunk space. FX has similar characteristics to DM---when the number of disks and the size of each dimension are a power of 2, FX is optimal for partial match queries. The scalability for range queries remains problematic.
\item \textit{Cyclic declustering.} The cyclic allocation scheme introduced in~\cite{divy:cyclic-declustering-2D} is a general declustering method for 2-D dense grids. Chunk $\left[i_{1}, i_{2}\right]$ is assigned to disk $\left(H * i_{1} + i_{2}\right) \mod \text{\textit{HD}} + 1$, where $H$ is chosen to be relatively prime with $\text{\textit{HD}}$. This results in separating close chunks in both dimensions on different disks---neighboring chunks on the same row are assigned to consecutive disks, while neighboring chunks on the same column are assigned to disks having distance $H$ apart. The condition that $H$ and $\text{\textit{HD}}$ are relatively prime guarantees that chunks are assigned to all the available disks before considering the same disk again. It is straightforward to remark that DM is an instantiation of the cyclic allocation scheme when $H = 1$. Given a value for $\text{\textit{HD}}$, it is possible to create an entire class of cyclic allocations that choose all the relatively prime values between 1 and $\text{\textit{HD}}$ for $H$---if $\text{\textit{HD}}$ is prime the number of classes is the largest. Not all of them provide the same performance though. Identifying the best value for $H$ requires a time-consuming exhaustive search. Even if the search space is drastically reduced, a close to optimal value for $H$ can be found with high probability. The scheme with the best performance proposed by the authors in~\cite{divy:cyclic-declustering-2D} that avoids the exhaustive search is based on Fibonacci numbers. Given a value for $\text{\textit{HD}}$, $H$ is chosen such that $H = F\left(F^{-1}\left(\text{\textit{HD}}\right) - 1\right)$, where $F(x)$ is the closed-form equation for the $x^{th}$ Fibonacci number obtained after solving the recursion---if $\text{\textit{HD}}$ is a Fibonacci number, $H$ is the previous Fibonacci number based on this equation. If $\text{\textit{HD}}$ is not a Fibonacci number and the resulting $H$ is not even relatively prime with $\text{\textit{HD}}$, $H$ is forced to be a relatively prime number with $\text{\textit{HD}}$ by finding the closest such number to the result obtained by the direct application of the equation.

There are two problems with this approach as presented in~\cite{divy:cyclic-declustering-2D}. First, it is limited to 2-D arrays. It is not clear how to generalize it to higher multi-dimensional spaces. And if the analysis still holds in higher dimensions. The second problem is the performance measure used in the paper. For a given size, all the queries across the entire space are considered and their error is averaged. Then, the errors are averaged again over all possible sizes. The problem is that the number of queries is considerably different across the sizes. And the maximum error is also highly dependent on the query size. As a result, the impact an individual query error has on the overall error is not uniform across all the query sizes. To be precise, the maximum error formula for a size (32, 32) 2-D array as proposed in~\cite{divy:cyclic-declustering-2D} is given below---for comparison, we also provide the formula that gives each query the same weight:
\begin{equation}\label{eq:divy:cyclic}
\frac {\sum_{i=2}^{32} {\frac {\sum_{j=1}^{(32-i+1)^{2}} {i^{2} / {\left\lceil\frac{i^{2}}{32}\right\rceil}}}  {(32-i+1)^{2}} } } {\sum_{i=2}^{32} }
\hspace{1cm}\text{ vs. }\hspace{1cm}
\frac {\sum_{i=2}^{32} {\sum_{j=1}^{(32-i+1)^{2}} {i^{2} / {\left\lceil\frac{i^{2}}{32}\right\rceil}}} } {\sum_{i=2}^{32} {\sum_{j=1}^{(32-i+1)^{2}} }}
\end{equation}
\end{itemize}

\paragraph{Space-filling curves.}
Another alternative to assign chunks to disks is based on the linearization provided by space-filling curves rather than by multi-dimensional chunk coordinates. A space filling curve visits all the points in a multi-dimensional space exactly once and never crosses itself. In this solution, chunks are first linearized using a space-filling curve that maps a multi-dimensional space into a linear sequence while preserving spatial proximity and then they are assigned to disks in round-robin fashion. Unlike cyclic declustering which enforces that neighboring chunks on both dimensions are spread apart as far as possible, space-filling curves guarantee this property only for a subset of the dimensions. Formally, chunk $\left[i_{1}, i_{2}, \dots, i_{N}\right]$ is assigned to disk $\text{\textit{Mapping}}_{\text{\textit{disk}}}\left(\text{\textit{Mapping}}_{\text{\textit{linear}}}\left(i_{1}, i_{2}, \dots, i_{N}\right)\right) \mod \text{\textit{HD}} + 1$, where $\text{\textit{Mapping}}_{\text{\textit{linear}}}$ is a space-filling curve---a complicated function at the border between implicit and explicit mappings. Out of the many space-filling curves proposed in the literature, the linearization based on Hilbert curves~\cite{faloutsos:hcam} is shown to provide the best performance both for partial match as well as range queries across multi-dimensional spaces when the number of disks is large.

\paragraph{Similarity-based graph-theoretic methods.}
The main idea behind the previously presented declustering methods is to make sure that neighboring chunks get assigned to different disks. This results in spreading the I/O throughput across many disks in the case of queries that select spatially close regions, thus improved execution time. The degree to which this goal is reached is a property of each method. The approach taken in the similarity-based methods presented in~\cite{moon:minimax,declustering:graph} is to formulate declustering as a graph partitioning problem. The graph is generated by creating a vertex for every chunk and creating an edge for every pair of chunks---complete graph. The edges are weighted by the probability that their adjacent vertices are accessed together by a query. Declustering corresponds to a multi-way partitioning of the graph. Since the goal is to minimize response time by maximizing parallelism in disk accesses, chunks -- vertices in the graph -- that are likely to be accessed together should be assigned to different disks---separate convex components in the graph. This problem is a variant of the well-known Max-Cut problem, which is known to be NP-complete. As a result, the similarity-based graph-theoretic methods for declustering are heuristic algorithms for the Max-Cut problem and its converse--the Min-Cut problem.

\paragraph{Recursive declustering.}
Instead of applying declustering to a full array, a different alternative is to partition the array into multiple sub-arrays and then apply declustering for each sub-array separately. The same or different declustering strategies can be applied for each sub-array. This approach is known as recursive declustering. The block-cyclic array partitioning strategy introduced in~\cite{arraystore} is a typical example of recursive declustering. It consists in splitting an array into regular blocks of chunks and declustering each block individually.

\subsection{Chunk Organization}\label{ssec:array:storage:chunk-organization}

Once array cell membership to chunk is determined, the next step is to organize the cells inside the chunk. Remember that the I/O unit is the chunk. Even if only one cell is needed for a given task, the whole chunk has to be read from disk into memory. While I/O is supposed to be the most time-consuming operation, memory access and CPU processing have their share. Thus, it is important to also consider optimization strategies for these operations.

The standard order in which cells are stored for dense grids is identical to the order in which chunks are linearized on disk---same order for dimensions. Other dimension ordering is possible. It is not clear what effect has on query response time. What is important for dense grids though is the storage reduction that can be obtained by discarding the indexes corresponding to array cells inside a chunk. This technique is known as \textit{dimension suppression}~\cite{scidb:ssdbm-11}. It reduces significantly the size of the chunk -- for highly-dimensional grids -- thus the amount of data that has to be read from disk. The only requirement for dimension suppression to be applicable is the existence of an implicit mapping function from an index to the corresponding cell inside the chunk---exactly the same idea as for chunk linearization.

In the case of sparse arrays, it is not that clear how to store cells inside a chunk. The simplest idea is to completely ignore any ordering and to process any query by scanning all the cells. This is perfectly reasonable since checking if a cell has to be included in the processing of a given query takes only a conditional \texttt{if} instruction. Given the purely relational format of sparse array data, any indexing technique -- including bitmap indexing -- is equally applicable. Based on dimensions or on the attributes. In particular, bitmap indexing along dimensions~\cite{sparse:bitmap-index} represents a secondary method to discard overlapping chunks for range queries. The only effect of any indexing technique is reducing the number of cells that have to be inspected at the cost of building the index. As mentioned before, the benefits are unclear.

In~\cite{sparse:bitmap-index}, the authors provide a complete overview on how to organize cells inside a chunk for sparse arrays. They analyze the storage requirement of each technique as a function of multiple parameters such as dimensionality, density, and size of the array cell. They also provide detailed analytical costs for the time it takes to answer point and range queries for each of the analyzed schemes. The storage schemes presented in the paper are given in the following:
\begin{itemize}
\item \textit{Index-value pairs}. This is the straightforward relational representation of sparse data. The order of the pairs inside the chunk can be arbitrary or it can follow the dimensions.
\item \textit{Offset-value pairs.} The same principle behind linearizing chunks on disk is applied to linearizing array cells inside the chunk. While absolute coordinates have to be stored for chunks, only the relative position in the chosen order is stored for array cells.
\item \textit{Compressed sparse dimensions.} In this representation, one of the dimensions is chosen as principal dimension. The position of each non-empty array cell is stored on this principal dimension in a 1-D vector. The cells are also stored in a corresponding 1-D vector. For the remaining dimensions, the transition from one index value to the next is recorded as positions in the 1-D vectors with indexes and array cells, respectively. While the size of the first two vectors depends only on the number of non-empty cells, the size of the last vector is equal to $\sum_{j=1}^{N-1}\left(\prod_{i=j}^{N-1}\left|D_{i}\right|\right)$, where $D_{N}$ is the principal dimension. Determining the order of the dimensions to minimize storage is quite straightforward. That is not the case for determining the principal dimension.
\item \textit{Sparse-dense split storage.} Dimensions are split into dense and sparse. When the dimensionality of the original array is reduced to the number of dense dimensions, the resulting arrays -- one for each combination of the sparse dimensions -- are either dense or empty. Empty arrays do not need to be stored at all. What has to be stored though are the indexes of the sparse dimensions for which there exist dense arrays.
\item \textit{Bit-encoded sparse storage.} Rather than storing the index of each dimension as a basic numeric type, e.g., \texttt{int}, \texttt{long}, the minimum number of bits sufficient to represent the cardinality of each dimension is used. This has the potential to result in storage reduction especially when chunking is used. Point queries also benefit from this representation since index matching becomes a bit manipulation operation---much faster than unoptimized integer multiplication and division. That is not the case for range queries which become more intricate.
\end{itemize}

Array cell organization inside a chunk can be viewed as another chunking problem, at lower scale. Thus, \textit{recursive chunking} can be applied. Everything discussed earlier applies directly to the more confined space. The depth of the recursion can be decided during chunking. In~\cite{arraystore}, the authors set for a two-level recursion. The benefit of such a strategy is again the reduction on the number of cells that are inspected in range queries. Notice though that the I/O unit remains the chunk at the upper-most level, independent of the number of levels.

%%%%%%%%%%%%%%%%%%%%%%%%%%%%%%%%%%%%%%%%%%%%%%%%%%%%%%%%%%%%%%%%%%%%%%%%%%%%%%%%%%
%\input{array-languages}
\section{Array Query Languages}\label{sec:query-lang}

Array processing is a common operation across multiple domains, including image processing and scientific computing. This results in a multitude of array operation types that have to be considered when designing an array query language. While several attempts have been made over the years, there is no commonly accepted array query language similar to SQL for relational data to date. The common trends among the proposed languages are to first identify an array algebra -- a set of primitive operators that can express as many array operations as possible -- and then to design a query language on top of the identified operators that resembles SQL as much as possible---typically array extensions to SQL. The biggest challenge faced when identifying the array algebra operators is the diversity of array operations mentioned previously. The standard solution is to allow for second-order operators -- operators that take user-defined functions as arguments -- in the algebra. Writing composite expressions of array algebra operator invocations is the first step in designing a query language. Several attempts stop at this stage. Adding a more elevated syntax on top of the pure algebra operator invocation is the next stage. To encourage adoption, the proposed syntax is quite often a modification to SQL---if not simple extensions with new keywords corresponding to the array algebra operators. In this more advanced scenario, query execution requires mapping the higher-level language constructs into array algebra operators---they are the only implemented functions that can be executed. If multiple mappings are possible -- the case when multiple implementations for the same operator are available or when the query expression permits it -- the optimal mapping has to be determined. This process corresponds to query optimization. In this section, we discuss multiple of the array query languages proposed in the literature in details.

\subsection{Map-Reduce}\label{ssec:query-lang:T2}

The first to remark the large variety of array operations were the authors of T2~\cite{t2}, the first full-fledged system for multi-dimensional data processing. Their goal was not to create an array query language though, but rather to design a parallel processing architecture that supports the most general form of array operations. As such, the approach taken in~\cite{t2} is to identify a standard form of array processing and then to build a parallel system that supports it. The standard processing supported by T2 takes as input a multi-dimensional grid and produces as output a different grid by applying the following functions: 1) transform input grid cells into items (transform phase); 2) map the transformed items to output grid cells (map phase); 3) aggregate all the input items mapped to the same output grid cell to compute the output value (reduce phase). Essentially, this is nothing else than Map-Reduce processing~\cite{google:mapreduce}. The defining characteristic of the parallel processing framework designed to support this type of processing is customization---each of the processing functions can be specified by the user at query time. If these are not available, the user is given the ability to implement its own functions satisfying a well-defined interface. Thus, while no query language is proposed, T2 provides a common framework to express a large variety of array operations using a common interface.

\subsection{Array Query Language (AQL)}\label{ssec:query-lang:AQL}

AQL~\cite{AQL} is a declarative query language for multi-dimensional arrays that treats arrays as functions from index sets to values rather than as collection types. This allows for expressing array operations in a higher-level language that hides the user from implementation details and is amenable to optimizations that would otherwise have to be implemented explicitly by the programmer. The negative side is a considerable reduction in the expressiveness of the operations that can be coded directly in the language. AQL addresses this drawback by providing extensible support for integrating user code dynamically in the language.

AQL is based on the nested relational calculus with arrays (NRCA) which plays the same role relational calculus and algebra play for the relational data model. Types and functions represent primitives in NRCA. The types include booleans, natural numbers, tuples, finite sets, and arrays defined over rectangular domains with indexes ranging over initial segments of the natural numbers. Functions are defined from one type to another. The constructs supported in NRCA not involving arrays are standard in nested relational calculus and include functions, products, set constructs, ordering, nesting, and arithmetic operators for natural numbers. There are exactly four array constructs:
\begin{itemize}
\item Define or tabulate an array
\item Extract an array element at a given index
\item Extract the dimensions of an array
\item Convert an indexed set into an array
\end{itemize}
These basic operators together with the standard constructs in the nested relational calculus are sufficient to express any operation on multi-dimensional arrays, including mapping a function to each element of an array, zipping multiple arrays together, i.e., positional natural join, extracting a subsequence -- not necessarily contiguous -- of an array, reversing, transposing, and projecting an array, and matrix multiplication. To simplify programming -- syntactic sugar -- and enhance query optimization, a series of derived constructs such as comprehensions, patterns, and blocks are also added as operators in the language---in a similar manner to operators in extended relational algebra.

An implementation of AQL in the ML functional programming language is introduced in~\cite{AQL}. AQL constructs are supported as library functions written in ML and made available as language operators. Queries are written as ML programs invoking these operators on the input data. Thus, there is no higher-level query language beyond the NRCA constructs. AQL simply takes advantage of the advanced programming features available in ML, including second-order functions. This is the main feature used to provide support for user-defined functions, thus extensible and customizable array processing. The operators are executed in full and intermediate results are materialized after each invocation. Part of query execution, the AQL constructs go through a series of transformations meant to generate an optimal execution plan that is eventually executed as calls to routines in the AQL library. Notice though that this type of optimization does not map a higher-level language into the AQL operators, but rather rewrites a sequence of function invocations optimally. I/O drivers that read/write data from/into various scientific storage formats allow for data to be loaded into AQL. It is important to notice that the proposed AQL implementation operates only on memory-resident data. There is no support for standard disk-based processing, storage management, and any other features characteristic to databases. Nor any discussion on optimal array representations. Thus, this implementation is only meant to showcase some simplistic array processing using the proposed AQL constructs.

\subsection{RasDaMan Query Language (RasQL)}\label{ssec:query-lang:rasql}

The RasDaMan~\cite{rasdaman,baumann:ngits} array algebra (read Section~\ref{ssec:rasdaman} for more details on the RasDaMan system) conceptualizes arrays as functions from rectangular domains to cell values, similar to AQL~\cite{AQL}. There are three core constructs in the algebra that can express every array operation when composed together~\cite{baumann:vldbj}. The execution of each of these constructs is iteration-based and is safe---does not require recursion. While user-defined functions can be integrated in the algebra, they are not fundamental. The authors advocate against their use due to the complications they introduce in query optimization. The three core array algebra constructs are the following:
\begin{itemize}
\item \textit{MARRAY.} The array constructor MARRAY creates new arrays by indicating a spatial domain and an expression which is evaluated for each cell position of the spatial domain. An iteration variable bound to a spatial domain is available in the cell expression so that a cell's value can depend on its position.
\item \textit{COND.} The condenser COND takes the values of an array's cells and combines them through the operation provided -- commutative and associative -- thereby obtaining a scalar value. An iterator variable is bound to the array spatial domain to address cell values in the condensing expression.
\item \textit{SORT.} The array sorter SORT proceeds along a selected dimension to reorder the corresponding hyper-slices. It rearranges a given array along a specified dimension based on an order-generating function which associates a sequential position to each (N-1)-dimensional hyper-slice without changing its value set or the spatial domain.
\end{itemize}
While these three operators are minimal to make the array algebra complete, a series of derived operators are added to the algebra to enhance usability. They are trimming and slicing, operators induced by the underlying type of the array cells, and multiple aggregates that are particular condenser instances. The result is an extended array algebra identical in spirit to the extended relational algebra.

Having the proposed array algebra as a theoretical foundation, RasQL is proposed as a declarative query language that extends SQL-92 with support for arrays. In RasQL, array expressions can appear in the SELECT and WHERE clauses of a SQL query. Special language constructs are introduced for the core array algebra operators -- MARRAY, COND, and SORT -- which can then be integrated with standard SQL. Following the SQL standard though, arrays are treated as a composite attribute type with a set of corresponding operators. Nonetheless, RasQL is the first complete array query language that integrates both an algebra as well as a higher-level declarative query language.

\subsection{Array Manipulation Language (AML)}\label{ssec:query-lang:AML}

The Array Manipulation Language (AML)~\cite{marathe:arrays} is an algebra consisting of three operators that manipulate dense arrays. Each operator takes one or more arrays as arguments and produces an array as result. All of the AML operators take bit patterns as parameters. Patterns are not allowed to refer to array element values. This restriction implies that the shape of the result of an AML operation can always be determined without actually evaluating the operator if the shapes of the operator's array arguments are known---the same is true for the schema of the result relation in relational algebra. This property is useful when evaluating AML expressions since it implies that the space required to implement an AML operation can be determined in advance. AML expressions can be treated declaratively and can be subjected to rewrite optimizations according to equivalence rules between operators.

The AML algebra operators are presented in the following:
\begin{itemize}
\item \textit{SUB.} Subsample is a unary operator that can delete data. The subsample operator takes an array, a dimension number, and a bit pattern as parameters, and produces an array, i.e., $B = \text{\textit{SUB}}_{i}(P, A)$, where A is the array, P is the bit pattern, and i is the dimension. SUB divides A into slabs along dimension i and then retains or discards the slabs based on pattern P. If P[k] = 1, then slab k is retained, otherwise it is not. The retained slabs are concatenated to produce the result array B. It can be shown that two subsequent SUB applications to two different dimensions of the same array produce the same result independent of their order, i.e., SUB is commutative across dimensions. This is not true when SUB is applied to the same dimension. Nonetheless, the resulting array can be inferred from the two bit patterns without the need to actually compute the result of each individual SUB operation. These rules are applied in query optimization.
\item \textit{MERGE.} Merge is a binary operator that combines two arrays defined over the same domain. The merge operator takes two arrays, a dimension number, a bit pattern, and a default value as parameters. It merges the two arrays to produce its result, i.e., $C = \text{\textit{MERGE}}_{i}(P, A, B, \delta)$, where A and B are the input arrays, P is the bit pattern, and $\delta$ is a default value. Conceptually, MERGE divides both A and B into slabs along dimension i. C is obtained by merging these slabs according to the pattern P. Because of shape mismatches between A and B, however, or because of the particular pattern P, some values in C may be undefined. $\delta$ is assigned to all such undefined values. It is important to remark that MERGE can be used to increase the dimensionality of an array. MERGE is commutative and associative when applied to the same dimension. Not with the same patterns though---the corresponding patterns can be easily determined. SUB and MERGE can be reordered both when applied to the same dimension as well as when applied to different dimensions. The corresponding patterns have to be determined from the patterns in the original expression. Choosing the optimal rule to apply is handled in query optimization.
\item \textit{APPLY.} Apply applies a user-defined function to an array to produce a new array. It is written as $B = \text{\textit{APPLY}}(f, A, P_{0}, P_{1}, \dots, P_{N-1})$, where f is the function to be applied, A is the array to apply it to, $P_{i}$'s are patterns, and N is A's dimensionality. APPLY makes the structural relationship explicit between array cells f is applied to through the patterns P. f is required to be defined such that it maps sub-arrays of A of some fixed shape $D_{f}$ to sub-arrays of B of some fixed shape $R_{f}$. APPLY applies f to some or all of the sub-arrays of shape $D_{f}$ of A. The pattern arguments specify to which of the possible sub-arrays of the input array A function f is applied. Pattern $P_{i}$ selects the slabs in dimension i. f is applied to the sub-array with origin at x only if x falls in selected slabs in all N array dimensions. Moreover, the sub-arrays to which f is applied to must be entirely contained within A. The results of these applications are concatenated to generate B. The arrangement of the resulting sub-arrays in B preserves the spatial arrangement of the selected sub-arrays in A. Applying a function to each cell of the array and to a chunk are special instances of APPLY. In~\cite{marathe:arrays}, the authors introduce two theorems showing how the structural locality captured by APPLY can be used to reduce the number of applications of f or to identify and eliminate unnecessary portions of the input array. When combined with the rewriting rules for the other operators, better execution plans can be determined.
\end{itemize}

AML -- as well as the RasDaMan array algebra -- is derived from an image algebra that defines the most common operations in image processing. The primary goal is different though. AML defines only those operators that are amenable to declarative optimization. Even so, a large class of image processing algorithms can be expressed in AML. With singleton APPLY's, i.e., APPLY is defined for each array cell individually, AML encompasses almost all the image processing algorithms. While image processing represents a large class of array manipulations, it is interesting to determine how AML handles other array operations that are not originating from image processing.  AML is a functional programming language in which operators are nested as arguments to other upper-level operators to form queries. Processing functions are also passed as functor arguments -- second-order functions -- to operators, i.e., the function argument to APPLY. Query optimization involves simple rewriting rules that replace combinations of algebra operators with other such combinations deemed optimal. Thus, AML is more like an elevated execution plan description rather than a declarative array query language. Another AML limitation is that it contains only structural operators, i.e., operators that consider the indexes. A complete description of the entire AML query evaluation process as implemented in ArrayDB~\cite{marathe:arrays} is presented in Section~\ref{ssec:arraydb}.

\subsection{Relational Array Mapping (RAM)}\label{ssec:query-lang:ram}

RAM~\cite{ram} is an array processing system built on top of the MonetDB~\cite{monetdb:overview} relational database. While RAM deals with dense arrays, SRAM~\cite{sram} is targeted at sparse arrays commonly used in information retrieval applications. Nonetheless, both systems employ similar array formalizations based on the comprehension syntax which represents arrays as functions defined over dimensions and taking primitive type values. Dimensions are defined over continuous integer intervals starting at $0$ for a regular array shape---not necessarily symmetric though. Array decomposition -- an array with composite type values is represented as a set of aligned arrays with primitive type values -- is default in RAM due to the columnar data representation in MonetDB. Since the execution happens inside a relational database engine, array queries follow a sequence of transformations that take arrays represented in the comprehension syntax to relational operators through an intermediate array algebra stage. Although a series of rewriting rules and optimizations are applied at each of these two steps, relying on the relational algebra operators to map and process array operations introduces inefficiencies due to the impedance mismatch in representation.

The RAM query language consists of methods to extract values from arrays and methods to construct arrays. Value extraction is supported natively through array application since arrays are functions that can be applied to index values to yield results. Array construction is supported through a generative comprehension constructor and a concatenation operator. There is no query language syntax defined for these functions, they are pure theoretic notations expressed in comprehension syntax.

The RAM array algebra consists of six operators that implement the query language---create arrays and extract values based on indexes only. The \textit{const} operator fills a new array with a constant value, whereas the \textit{grid} operator creates an array with values taken from one of the indexes. \textit{map}, \textit{apply}, and \textit{choice} are induced operators that operate on cell values. \textit{map} creates a new array by applying a given function to one or multiple aligned arrays. \textit{apply} replaces the function in \textit{map} with an array interpreted as a function from indexes to values. \textit{choice} is a combination of \textit{map} and \textit{apply} in which an array with boolean values selects the elements of a newly created array from the elements of two arrays passed as arguments. And, finally, the \textit{aggregate} operator applies an aggregate function passed as argument to the array elements having the same value for the first $k$ indexes, resulting in an array with smaller dimensionality.

In addition to the RAM operators, the SRAM array algebra~\cite{sram} introduces a series of additional structural operators which are missing entirely from the RAM specification. \textit{pivot} permutes the dimensions of an array according to an axis order permutation. \textit{rangeSel} is the standard subsample operator which extracts a sub-array with the same dimensionality from an array passed as argument. \textit{replicate} generates an array with dimensionality $N+1$ by replicating the original array a specified number of times. \textit{topN} is a very specialized operator that works only for vectors and creates an array with the indexes of the first $K$ values in a specified order.

The mapping of the extended SRAM array algebra operators to relational operators is presented in~\cite{sram}. It is very specific to the chosen relational representation of arrays in MonetDB. Sparse arrays are stored as relations clustered and indexed based on the array dimensions. The order is chosen arbitrarily as the lexicographical dimension order, i.e., the order in which dimensions are specified in the array definition. Only the cells with valid values are stored explicitly. The mapping of \textit{apply} as a series of joins followed by a projection is presented as a canonical mapping for all the structural operators---\textit{pivot}, \textit{rangeSel}, and \textit{replicate}. \textit{map} between two dense arrays corresponds to relational join followed by function application. In the case of sparse arrays, the general form of outer join is used instead. \textit{aggregate} can be mapped into a standard group-by aggregate relational operator on the dimensions, while \textit{topN} does not have a relational equivalent. In addition to the mapping rules from array algebra operators to relational operators, a series of trivial optimization and arithmetic simplification rules are also introduced. All the presented rules prove that while the process is possible, it is also very complicated. Even if it is entirely automated inside an optimization stage, we have serious doubts on its efficiency, especially when compared to a dedicated implementation that preserves the data management functionality of a database server, for example, implement the array algebra operators as user-defined functions (UDF) and user-defined aggregates (UDA).

\subsection{SciQL}\label{ssec:query-lang:sciql}

SciQL~\cite{SciQL-edbt,SciQL} is the most comprehensive extension to the SQL:2003 standard with support for arrays. It provides seamless integration of set, sequence, and array semantics. The goal is to make minimal modifications to the SQL syntax while allowing for maximum expressiveness in the array operations supported by the language. It is heavily targeted at experienced SQL programmers. While this is considered to be one of the most distinctive characteristics of SciQL from a database perspective, it might also be an important drawback given the reduced familiarity the science community has with SQL.

SciQL provides all the benefits of a declarative query language that isolates an abstract data model from the physical data representation. Arrays are defined by specifying the dimensions, their corresponding ranges, and the array cell content. Named dimensions allow for direct indexing of the array elements. A default value is assigned to all the cells in the array at declaration. Arrays can appear wherever tables are allowed in an SQL expression. The result of a query is an array only when the column list of a SELECT statement contains dimensional expressions. The SQL iterator semantics associated with tables extends to arrays, but iteration is confined to the cells whose values are not NULL. This might be quite inefficient though for operations that require array traversal in a particular order.

Array creation and modification statements follow entirely the syntax corresponding to tables. The only difference is that dimensions have to be defined explicitly for arrays. Converting arrays to tables can be done by simply selecting all the array cells without specifying any dimensional expression. The reverse is not that straightforward since the designated dimensions have to form a primary key in both representations. If the result of a query is an array, it has to be specified explicitly in the SELECT clause. Cell selection and array slicing are performed using the bracketed index syntax from \texttt{C}. The most specific array operator introduced in SciQL is structural grouping---in fact, it is a syntactic representation for the \textit{APPLY} operator introduced in~\cite{marathe:arrays}. It consists in placing a template at every position in the array and computing an aggregate for all the cells in the neighborhood that are covered by the template. The result is an array with the same dimensions. Two versions are proposed---with and without overlap. SciQL provides extensibility through user-defined operators. They can be implemented using primitive SciQL constructs -- similar to stored procedures -- or they can be imported from an imperative programming language like \texttt{C}---similar to UDFs in standard databases. In addition to multi-dimensional array operations, SciQL supports a large range of time-series operators. We do not discuss these features in the paper since they are out of the main scope.

To evaluate the functionality of the language, a series of examples from different scientific domains are provided. The purpose is to show how complex scientific processing can be expressed in SciQL. While definitely possible, it is interesting to examine what features of the language are used and what is the complexity of the SciQL constructs. It is no surprise that almost the entire processing is done through UDFs---the examples first define one or more UDFs that are subsequently invoked in the SELECT and WHERE clauses. This is due to the complexity of the processing which rarely maps well on the restrictions imposed by a strict declarative language. Structural grouping is the specific array operator mostly used throughout the examples. It can be seen as a UDF promoted to the rank of an operator with assigned syntax. The expected simplification in complexity specific to a declarative language is not always possible for scientific processing. Some of the SciQL statements provided as examples are as large as half a text column. This casts serious doubts on the expressiveness of SciQL. The most important concern we have is the following:
\begin{itemize}
\item Why would science adopt SciQL when they are not even using SQL in the first place?
\end{itemize}
Astronomy is an exception, but even there SQL is mostly used for metadata and derived data querying.

The last point we want to discuss on SciQL is the implementation of the language. There is no array algebra to support the language. And no implementation details are provided at the time when we write these lines. Since SciQL is the work of the same research group that introduced RAM~\cite{ram} and SRAM~\cite{sram} where an array algebra is mapped onto the MonetDB columnar relational algebra, it is interesting to see what alternative will be chosen this time.

\subsection{SciDB Query Languages}\label{ssec:query-lang:SciDB}

SciDB~\cite{scidb:ssdbm-11} is a shared-nothing parallel database system designed specifically for dense array processing (see Section~\ref{ssec:scidb} for more details). SciDB queries can be written in two languages---Array Functional Language (AFL) and Array Query Language (AQL). AFL is a functional language in which the execution plan is expressed exactly in the same format as in AML~\cite{marathe:arrays}. A slight difference is that the number of operators is larger than in AML. The reason is that instances of APPLY that execute a specific operation are promoted to stand-alone operators with their own name. AQL is SQL adapted to array processing. From the examples provided, it looks like array operations are expressed as relational algebra operators. It is not clear how specific array manipulations coded as UDFs are expressed in AQL. Thus, given the released SciDB versions which use mostly AFL, we believe this is the main query language in SciDB at the time when this document was written.

\subsection{Discussion}\label{ssec:query-lang:discussion}

We end this section with a comparison of four array models -- AQL~\cite{AQL}, RasQL~\cite{baumann:vldbj,rasdaman,baumann:ngits}, AML~\cite{marathe:arrays}, and (S)RAM~\cite{ram,sram} -- based on the work in~\cite{baumann:compare-array-algebra}. Arrays are always modeled as functions from hyper-rectangles to primitive or composite values. Array creation is specified using either tabulation (RasQL) or comprehension (AQL and RAM). Operations are defined as functionals, i.e., second-order functions taking other functions as parameters. While this generates a small set of operators, a large part of the complexity is hidden in the functional parameters. An important question that has to be answered is how many physical operators to implement and to make available through the language syntax? The answer can vary from all -- the case in SciDB -- to only the operators in the algebra. There is no definitive answer though.

In~\cite{baumann:compare-array-algebra}, it is shown that all the array algebras can be reduced to RasQL---both in array representation as well as operations. This is primarily due to the equivalence between comprehensions and the \textit{MARRAY} operator for creating arrays---the comprehension syntax is the basis for AQL and RAM. The equivalence between AML and RasQL is proven directly---it is valid in both directions. Extensive examples showing how a large variety of array operations are expressed in RasQL are presented in~\cite{baumann:vldbj,baumann:ngits}. Given the reduction to RasQL, all the examples shown for the other algebras can be immediately mapped to RasQL expressions.

There are other array models proposed in the literature that we do not discuss in detail. AQUERY~\cite{aquery} uses the concept of arrables, i.e., ordered relational tables and SQL queries extended with an ASSUMING ORDER clause to represent one-dimensional time series data. Howe and Maier~\cite{howe:array-algebra} propose a blob-based approach where an algebra for the manipulation of irregular topological structures is applied to the natural science domain.

%%%%%%%%%%%%%%%%%%%%%%%%%%%%%%%%%%%%%%%%%%%%%%%%%%%%%%%%%%%%%%%%%%%%%%%%%%%%%%%%%%
%\input{array-dbms}
\section{Array Processing Systems}\label{sec:array-syst}

When it comes to handling large-scale arrays that require out-of-memory processing, techniques from databases have to be applied. As with any other system targeted at a different data model than the original it was built around, there are two alternatives. The first alternative is to map the array data model onto the existing data model. In the case of databases, this model is the relational data model. Mapping arrays onto relations requires two steps---representation mapping and operation mapping. Representation mapping considers how to map the array structure over relations, while operation mapping consists in expressing array processing operations as relational algebra operators. While generally possible, the problem with this approach is that the mapping is rarely perfect and typically requires convoluted transformations that result in unacceptable performance degradation. The advantage is that the starting point is an existing system that has all the basic functionality implemented and thoroughly tested. The alternative is to design and build a system targeted to array processing that implements specialized optimizations from scratch. While the amount of work is considerably more strenuous so are the potential performance benefits. In the following, we present more details on how arrays are implemented inside a relational database systems. Then, we move to specialized systems for array processing and present the details of such implementations for multiple prototype systems proposed in the literature.

\subsection{Arrays in Relational Database Systems}\label{ssec:array-syst:rel}

There are multiple approaches to support arrays in relational database management systems:
\begin{itemize}
\item \textit{Relational representation.} Each array cell is represented explicitly as a tuple containing both the indexes as well as the values. Array manipulations are then mapped into relational algebra operators and are expressed using standard SQL. However, SQL is not particularly well-suited for array operations due to the limited expressiveness targeted at relational algebra operators. To compensate for the lack of expressiveness in the language, array manipulations can be coded directly as user-defined functions. This requires though out-and-back conversion from the relational representation to arrays which incurs additional execution time.
\item \textit{Array composite data type.} In this approach, specific to object-relational database systems such as PostgreSQL~\cite{postgresql}, a composite array data type and corresponding operators are supported natively. Array operations can then be included in queries by making calls to the methods defined for the array type. Each query consists of a relational component and a non-relational component, where the non-relational part contains expressions involving array methods. The problem is that, in most object-relational database systems, optimization of the non-relational parts is very limited, with method invocations being treated as black boxes. At best, the optimizer is capable of handling only the placement of the non-relational parts of the query within the relational execution plan.
\item \textit{BLOB storage.} Arrays are represented as binary large objects (BLOB). The system provides only array storage but does not support array manipulations. In the best case, it is possible to select a portion of the array by retrieving only the corresponding portion of the BLOB. All other array manipulations have to be implemented by the application itself though.
\end{itemize}
The standard relational representation approach is directly supported by any relational database system. Complex array manipulations on relational data require support for user-defined functions (UDF)---a common feature in any modern database. When functions are coupled with a specific data type, as is the case in object-oriented programming, and new types and corresponding methods can be added to the system in the form of user-defined data types (UDT), the requirements to support arrays as a composite data type are satisfied. While this is mostly specific to object-relational databases, almost any modern database system supports arrays as a composite type. The same holds for BLOBs, the standard representation for large objects such as image and video files.

\subsection{Array Database Systems}\label{ssec:array-syst:dbms}

Array database systems are often designed for specific application domains such as scientific computing and online analytical processing (OLAP). Several such systems were proposed in the literature over time. In the following, we review the most important systems and their defining characteristics.

\subsubsection{Titan}\label{ssec:titan}

Titan~\cite{titan} is a parallel shared-nothing database designed for handling remote-sensing raw data obtained from satellites in the form of AVHRR files. Titan architecture is standard for parallel databases with one node acting as the coordinator and the other processing nodes being workers. The coordinator receives queries from clients and schedules query processing across the workers. The coordinator stores only metadata in the form of an R-tree index built over the data chunks and kept in memory due to its reduced size. The workers are assigned data chunks that are stored on their local disks. Queries are specified as spatio-temporal regions and require mapping multiple points in the input data to a single point on the output grid---typical \textit{group-by aggregation} query in relational databases. Given this simplified processing, query execution follows a standard pattern. The output grid is split into multiple regular chunks, one of each being assigned to each worker node for processing. The input chunks that intersect the query range are identified using the metadata index at the coordinator, scheduled for reading by the workers that store them, and dispatched to all the corresponding workers that need them for processing. Each worker node is responsible for scheduling the order in which to read chunks from the local disks. Fortunately, the order in which chunks are processed does not matter since all the aggregates are associative decomposable. Thus, the schedule has to balance only between optimal reading from disk and providing chunks to all the nodes that need them in a fair fashion. No details are provided on how this is achieved.

The main processing loop at each worker node consists of five asynchronous stages executed in the following order:
\begin{enumerate}
\item Chunk read request from local disks. As many asynchronous chunk reads as possible are requested from the local disks according to the determined schedule. The limiting factors are the available memory and the number of concurrent requests that can be made to the same disk.
\item Chunk send to other workers. The buffers containing read chunks that finished transfer to processing nodes are freed for subsequent usage.
\item Chunk receive from other workers. Requests are made to other worker nodes to produce chunks needed for processing locally. The number of such requests is also maximized based on memory availability and the number of concurrent requests from the same worker. The chunks received in full are prepared for processing. This stage is a bit confusing since workers generate chunks based on their local schedule. If a chunk read into memory is not immediately sent to all its consumers, it occupies resources indefinitely. Chunks are produced according to the producer's schedule and consumed according to the consumer's availability---a mixture of push- and pull-based processing. This disconnect is a serious source of sub-optimal performance in memory usage and it is prone to potential deadlock. The authors do not mention any of these problems.
\item Chunk read check from local disks. The chunks that are done reading from disk are either prepared for local processing or scheduled for transmission to remote worker nodes.
\item Chunk process. If there is any available chunk for processing, the composition operation is executed. Since processing is part of the same loop as all the previous operations, at most one chunk is processed at a time to allow query execution progress.
\end{enumerate}
Query execution is not presented well. What happens is more likely to be push-oriented processing driven by the reading schedule. When chunks are read from disk, consumer workers are signaled. The transfer does not start immediately though. It is originated only when the consumer is ready to receive the chunk. Meanwhile, the chunk sits in the producer's memory and this might have negative effects on query execution, including complete stall in the worst case. Instead of processing only one chunk per iteration, processing can also be detached and run in an asynchronous process. The lack of multi-threading at the time might be the cause this was not considered as an alternative.

The input data are a series of 2-D dense grids ordered by time. Each array cell contains five attributes. The output data are 2-D images -- also dense grids -- generated from the composition of multiple 2-D input grids obtained at different time instants. Each grid in the input data is partitioned separately into aligned chunks as close to regular as possible. The chunk size -- 200-300 KB -- is chosen such that the I/O throughput is optimized for the specific machine Titan was designed for---Titan was built to work on a specialized parallel shared-nothing machine. Domain knowledge is used to decompose the original 5-attribute array into two arrays with 2 and 3 attributes, respectively, that are always queried separately---there is no need to join data across arrays in any way. The chunks corresponding to each of the two arrays are stored contiguously on disk. Chunk assignment to disks -- declustering -- is done using the \textit{minimax} graph-based algorithm~\cite{moon:minimax} proposed by the same authors. The minimax algorithm guarantees a perfect balanced distribution of chunks to disks at a quadratic cost in the number of chunks. The chunks assigned to the same disk are optimally clustered based on the \textit{Short Spanning Path (SSP)} algorithm which the authors showed to perform better than the Hilbert-curve-based algorithm~\cite{moon:minimax}. A single simplified R-tree index containing spatio-temporal information is created for all the chunks in all the input grids. Data needed to retrieve chunks based on their spatio-temporal position -- [disk, offset] pair -- is stored in the leaves of the index. The interior nodes contain bounding quadrilaterals -- not rectangles -- for their children. To allow for fast range queries, chunks are first sorted spatially in a z-ordering before the index is created.  

The main problem with Titan is that there is no connection between input data location and output processing---the scheduling of the output chunks to workers for processing is done without considering the location of the input data at all. This results in massive amounts of data that have to be transferred across the network, thus degraded performance. A possible solution is to execute partial aggregation locally at each node holding the data and passing only the intermediate result for final aggregation at the node storing the corresponding result chunk. This solution corresponds to combiners in Map-Reduce~\cite{google:mapreduce} and to local aggregation in GLADE~\cite{glade:osr}. While optimal declustering increases processing parallelism, it also spreads data too well across nodes. This results very likely in having all the nodes involved in the processing of every query. If this is the optimal processing strategy is questionable especially with the current multi-core architectures that support extensible parallelism at the level of a single node. The solution proposed by the authors for this problem is to increase the chunk size such that a considerable smaller number of chunks is required for processing, thus increasing locality. Since these problems show up even in the case of range queries -- the only experiments presented in the paper are range queries -- it is clear that in the case of clustering where data from neighboring chunks have to always be considered these problems are even more stringent. In conclusion, Titan seems to be the precursor of Map-Reduce~\cite{google:mapreduce} since the same processing strategy is used.

\subsubsection{T2}\label{ssec:t2}

T2~\cite{t2} is the first customizable parallel database that integrates storage, retrieval, and processing of multi-dimensional datasets built as an extension to Titan~\cite{titan}. Previous systems for multi-dimensional data management support only storage while unloading the entire processing to the applications due to their too specific nature. The standard processing supported by T2 takes as input a multi-dimensional grid and produces as output a different grid by applying the Map-Reduce processing paradigm introduced in Section~\ref{ssec:query-lang:T2}: 1) transform input grid cells into items; 2) map the transformed items to output grid cells; 3) aggregate all the input items mapped to the same output grid cell to compute the output value. Essentially, this is nothing else than \textit{group-by aggregation} in relational databases as long as each input cell can be mapped to a single output cell. Each of the processing functions can be specified by the user at query time, thus the customizable nature of T2. Additionally, T2 manages the allocation and scheduling of all resources in the parallel environment based on annotations provided by the user for each of the processing functions. The only supported processing in T2 consists in the user specifying the (region of interest in the) input dataset, the format and resolution of the output, and selecting the processing functions. If these are not available, the user is given the ability to provide its own processing functions satisfying a well-defined interface---T2 is also extensible.

The T2 architecture is modular and consists of a series of customizable services that have to be integrated together to support query processing. The attribute space service manages the use and registration of attribute spaces and of mapping functions between attribute spaces. The data loading service manages the loading of new datasets into T2. It takes as input already chunked data -- the chunking has to be performed by an external application -- and executes only the loading part according to two user-specified algorithms for declustering of chunks across multiple disks and for ordering the chunks assigned to the same disk, respectively. Both algorithms are customizable by the user at loading time. The data loading service computes a bounding rectangle for each chunk as derived data during the loading process. This is subsequently used together with the physical location on disks to create an index for a given [dataset, attribute space] pair by the indexing service. The data aggregation service is in charge of the processing functions---transformation and aggregation. The only supported aggregation functions are associative decomposable, i.e., commutative and associative, both distributive and algebraic. This allows for processing input grid cells in parallel and in any order. Users formulate queries in T2 using the query interface service. Queries are then scheduled for execution by the query planning service which manages the available resources. Two execution strategies are considered---pure Map-Reduce and Map-Reduce with combiner~\cite{google:mapreduce}. The final choice is made based on several factors such as output data distribution, input chunk placement, and machine physical characteristics. The query execution service executes the selected plan by applying the processing functions in the correct order to all the items in every chunk. Several optimizations are applied to enhance the query execution performance. Chunks are read only once from disk and all the required processing is applied when this happens. Chunks are never copied across the different services. All the processing functions are executed at the storage manager level the first time the chunk hits the memory. Disk I/O, network operations, and processing are overlapped as much as possible in a similar manner to the Titan~\cite{titan} main processing loop.

The authors present the implementation of Titan~\cite{titan} in T2 as a concrete customization example. Essentially, all the processing performed in Titan is implemented as processing functions in T2. These are then put together as a query specification and passed to T2 for execution. As such, T2 is nothing else than a generalization of Titan. Unfortunately, no other examples beyond Titan are presented. To extend the similarities, T2 can be viewed as the direct precursor of the Map-Reduce framework~\cite{google:mapreduce} including all the customization and extensibility features that lacked in Titan.

\subsubsection{RasDaMan}\label{ssec:rasdaman}

RasDaMan~\cite{rasdaman} is a domain-independent array DBMS with support for multi-dimensional arrays of arbitrary size, dimension, and structure through a general-purpose declarative query language paired with internal execution, storage, and transfer optimization. RasDaMan has a standard client-server architecture---the server runs on a single machine, i.e., not a parallel database. The server is a middleware that runs on top of a standard (relational) database with support for BLOB storage---chunks are stored as BLOBs. The arbitrary chunking strategies supported in RasDaMan are presented in~\cite{furtado:tiling}. Array operations are executed entirely at the middleware level. The back-end database provides only storage support for chunks as BLOBs. The RasDaMan middleware replicates all the components of a standard relational DBMS. The query parser transforms a SQL-like query with array constructs -- a RasQL query (Section~\ref{ssec:query-lang:rasql}) -- into an operator-based query tree containing only operators defined in the array algebra---similar to relational algebra. The query optimizer transforms the query tree into a more efficient execution plan based on algebraic query rewriting rules and chunk layout information. The goal is to find the optimal order in which to access the chunks on disk. Once the optimal plan is determined and the chunks that have to be processed are identified based on the multi-dimensional index stored as part of the metadata catalog, the array operators are invoked through function calls. While not specified, the operators in the array algebra take as input array(s) and generate array(s), i.e., only arrays flow between operators. Extensive implementation details for chunk processing in RasDaMan are presented in~\cite{rasdaman:execution}.

\subsubsection{ArrayDB}\label{ssec:arraydb}

ArrayDB~\cite{marathe:arrays} is a prototype array database system that implements the AML language (see Section~\ref{ssec:query-lang:AML} for details) in which arbitrary externally-defined functions can be applied to arrays in a structured manner. AML execution plans pipeline data through operators and generate results a piece at a time by choosing the optimal result generation order. There are five steps in the processing of AML queries. They take a query from the AML expression to the execution tree consisting of the array algebra operators presented in Section~\ref{ssec:query-lang:AML} and then feed them with data in the input arrays to produce the result array. This process is similar to how relational algebra expressions are evaluated in relational databases.

The \textit{pre-processor} generates the initial query tree containing an internal node for each SUB, MERGE, and APPLY operator in the query, and a leaf node for each input array. Leaf nodes are treated as special chunked APPLY operators that generate the input array data a chunk at a time. The bit pattern can be used as a selection predicate for the chunks to be read. Based on metadata from catalogs, the pre-processor tags the nodes in the tree with dimensionality and schema information. It also introduces new MERGE operators to ensure that all the existing MERGE operators in the tree are balanced---the array operands have the proper dimensionality.

In the \textit{logical rewriting phase}, the query tree built by the pre-processor is transformed into an equivalent tree that is more efficient to evaluate. The cost measure is the number of applications of user-defined functions in the APPLY operators in the tree. The logical rewriting procedure finds a query tree with the minimum number of such function calls among the trees that are both equivalent to and apply-consistent with the original query tree---these conditions are meant to reduce the search space. It proceeds in N -- the highest array dimensionality in the query -- top-down traversals of the tree, where only operators on the $i^{\text{th}}$ dimension are considered at iteration i.

The optimized query tree is then mapped into a \textit{physical query execution plan} consisting of a series of six types of physical operators---APPLY, REPLICATE, REGROUP, COMBINE, LEAF, and REORDER. These physical operators are designed according to the pull-based iterator interface. They take as input chunks of a given shape from one or multiple arrays and produce as output a single chunk for each call to \texttt{NextChunk}. While it is desirable to have a completely pipelined plan in which a chunk can follow a full path in the tree continuously from the leaf to the root, this is not always possible due to the blocking nature of some of the operators---REPLICATE, REGROUP, and REORDER are blocking operators. Blocking operators require chunks to be buffered in memory before being processed---in the worst case, an entire array has to be buffered in memory. This is the case for the REORDER operator which restructures the chunks of an array. Since in a query execution plan successive operators must have compatible chunk shapes and chunk generation orders, multiple of the physical operators accomplish this task---the difference in number between the logical operators and the physical operators originates from this rather than having multiple physical implementations of the same logical operator, as is the case in relational databases. The simplifying assumption at the core of the entire execution strategy is that arrays are stored regularly chunked into same shape, same size chunks~\cite{sarawagi:chunking,furtado:tiling}. The chunks are read sequentially from disk and delivered into the system for processing as part of the \texttt{NextChunk} calls propagated from the root to the leaves.

A second round of \textit{query optimization} is performed on the physical execution plan. The cost measure is in this case the memory buffer space taken by the plan operators. Memory cost reduction can be obtained in two ways. First, unnecessary operators can be eliminated from the plan altogether. Second, the order in which each plan operator produces and consumes array chunks is chosen to be optimal across the overall plan. Whenever two successive operators have different orders, a REORDER operator has to be inserted that transforms the output order of the child operator into the input order of the parent. The question is where to place these operators optimally in a plan such that the overall memory usage is minimized. A bottom-up dynamic programming algorithm that solves this problem is presented in~\cite{marathe:arrays}. As soon as the optimal physical plan is determined, chunks can be read from the storage and \textit{query execution} can proceed.

ArrayDB supports structured execution of any user-defined functions over arrays. In theory, this allows for any type of array processing to be executed. In practice, ArrayDB is mostly used for processing small 2-D images. Little, to none, emphasis is put on array storage. The simplistic regular chunking strategy is assumed, with chunks stored on disk in row or column major. This is justified to some extent by the limited application to image processing. What is not justified though, are the cost measures used in optimization -- number of user-defined function applications and memory usage -- since none of them are standard optimization measures in databases. At the end of the day, ArrayDB is supposed to be a database not an in-memory system. Moreover, the two-step optimization strategy in which the number of function applications is minimized first, followed by minimizing memory usage, is not guaranteed to produce the overall optimal solution since each of them are local optimizations. It would be interesting to extend some of the ArrayDB ideas into a full-fledged \textit{real} array database system that handles the contemporary massive arrays. Parallelization is one of the first issues that need to be addressed since it is almost entirely ignored in ArrayDB---except for the pipelined execution. How does AML handle sparse array data? Or arrays with more than a single value per cell? How to extend the in-memory physical operators to disk-based operators? How to optimize queries in this case? These are just a few questions that require careful consideration.

\subsubsection{(S)RAM}\label{ssec:sram}

(S)RAM~\cite{ram,sram} implements the (S)RAM array algebra on top of the MonetDB/X100~\cite{monetdb:overview,monetdbX100} relational database system. As a result, the array algebra operators are mapped into relational algebra expressions containing selection, join, group by, and other relational algebra operators. The fundamental question is how to do the mapping. The execution is purely relational. The benefit of such an approach is that an existing system -- MonetDB in this case -- is used for processing. No system has to be rebuilt from scratch. The disadvantage is that the mapping is not always optimal due to the impedance mismatch between relations and arrays.

RAM is further extended to a parallel setting in~\cite{ram-distributed}. Two simple rules that allow array-specific query decomposition are presented---partitioning and aggregation. Partitioning allows for sub-arrays of the same array to be evaluated in parallel whenever there is no dependency between indexes. Aggregation allows for commutative and associative functions over arrays to be evaluated in parallel. These rules are straightforward extensions of the relational set semantics.

\subsubsection{SciDB}\label{ssec:scidb}

SciDB~\cite{scidb:ssdbm-11} is a shared-nothing parallel database system designed specifically for dense array processing. It supports multi-dimensional, nested arrays with array cells containing records, which in turn can contain multi-dimensional arrays. Every cell has the same data type(s) for its value(s), which are one or more scalar values, and/or one or more arrays. Sparse arrays are treated as dense arrays enhanced with a bitmap indicating which cells contain valid data. Arrays can be enhanced with UDFs to map a given coordinate system into another and to provide the shape of an irregular array. UDAs and table functions -- functions that take tables as arguments -- are also supported.

SciDB is novel in that it has no built-in operators. Instead, all operators are UDFs. Some are provided with the system and the user is given the freedom to implement any other operators using the extension facility provided by UDFs. The operators can be either structural, e.g., slice, subsample, reshape, concatenate, and cross-product; value-based, e.g., filter, aggregate, apply, and project; or both, e.g., join. As is the case in ArrayDB, a query execution plan consists of a series of successive UDF operators. The difference is that in SciDB all these operators are instances of APPLY, possibly with different processing functions. Unless successive UDFs are commutative, the structure of the query execution plan cannot be altered. Thus, query optimization focuses on parallelizing individual UDF operators and pipelining array chunks between operators---the I/O, network transfer, and processing unit is the chunk. The main idea behind query execution is to identify segments of successive operators that can be executed in pipelined fashion on a single node. They are scheduled and executed on the nodes where data are stored. Whenever data have to be transferred across nodes, chunks are moved in a coordinated process. After each such step, intermediate result information is gathered and used for the optimization and scheduling of the next segment---optimization and scheduling are executed at run-time.

SciDB introduces multiple contributions at storage level---array decomposition, chunk overlapping, and heavy use of chunk compression. The main idea is hardly new though---chunk the array into fixed size (logical) chunks across all dimensions, i.e., regular chunking. The size of the chunk is large enough to mask the seek time by the data returned. In the case of sparse arrays, where fixed size chunks can contain considerably fewer data, adjacent chunks can be merged together into a single chunk. Array decomposition consists in splitting an array with multiple values in a cell into multiple independent arrays with a single value in each cell. This is a generalization of the column-stores ideas to arrays. Chunk overlapping consists in storing the same array cells in multiple chunks to increase the level of parallelism in execution. Immediate drawbacks of this include increased storage and more complex management. Several compression techniques are extended from column-stores, including Null Suppression, Run-Length Encoding, subtraction from an average value, and Delta Encoding. They are applied on a chunk-by-chunk basis, with different chunks from the same array possibly compressed differently.

Extensibility in SciDB follows the patterns introduced in PostgreSQL~\cite{postgresql}. A user can add new types to the system or User-Defined Types (UDT); scalar functions taking arguments basic and user-defined types and returning a single value, i.e., User-Defined Functions (UDF); User-Defined Aggregates (UDA) which allow special aggregate computation for the newly defined types expressed as a group of functions invoked according to a well-established template; User-Defined Operators (UDO) taking arrays as arguments and producing an array as the result. While it is quite straightforward how to parallelize UDAs, implementing parallel UDOs is a considerably more difficult task, especially when the logic inside the operator consists of multiple complex steps with conditional logic between them. It is important to emphasize that UDAs are considered only as a mechanism to implement aggregate computations for UDTs. They are never considered a general processing paradigm capable of expressing almost any type of computation, the case in GLADE~\cite{glade:demo}.

SciDB is designed to be considerably more than an array DBMS. It is supposed to be a complete system for scientific data management and analysis. As such, it provides an entire suite of features targeted at scientific data in general. They include a complex versioning system with no data overwrite, in-situ data processing without previous loading into SciDB, array and cell provenance, and support for uncertain cell values. How exactly are all these features supported is unclear from the first SciDB papers which provide an overview of the system.

\subsubsection{ArrayStore}\label{ssec:arraystore}

ArrayStore~\cite{arraystore} is a storage manager for parallel array processing. ArrayStore is designed to support complex and varied operations on sparse arrays and parallel processing of these operations in a shared-nothing cluster. The workload supported by ArrayStore includes the following classes of array operations:
\begin{itemize}
\item \textit{Value-based operations.} These operations operate on the content of array cells, independent of their position. They are derived from the corresponding relational operations. Filter or selection is the representative operation in this class.
\item \textit{Structural operations.} Subsample is the representative structural operation. It selects a compact sub-part of the array based on the position of the cells. Only the cell index is important, not the content.
\item \textit{Binary operations.} These are structural extensions of relational join, i.e., a natural join between two arrays on cell indexes. The main question is how to partition the arrays across multiple sites to optimize the processing. The answer, known for a long time in parallel databases, is to chunk the arrays using the same strategy and co-locate corresponding chunks at the same site.
\item \textit{Overlap operations.} Access to neighboring array cells is required for each cell in the array. Problems arise for the cells on chunk borders since some of their neighbors reside in a different chunk, stored at a remote site, thus requiring data transfer. While parallel processing of chunks is possible, some sort of post-processing is required for this corner case of border cells.
\end{itemize}

The chunking strategies considered by ArrayStore for the varied workload described above are regular and irregular chunking, and a two-level combination of the two. Regular chunking is a well-known strategy studied in depth before~\cite{sarawagi:chunking,furtado:tiling}. Irregular chunking aims at creating chunks with the same number of points in each chunk rather than generating chunks with the same shape. The motivation is the desire to have uniform processing time across chunks in a parallel environment, thus load balancing. It is questionable if this goal is achieved by having chunks with the same size. Moreover, creating same size chunks is a complicated process that increases considerably the data loading time. Irregular chunking also requires explicit indexing for efficient chunk retrieval. Based on these elementary chunking strategies, a two-level chunking strategy that combines the two is advocated. The idea is to first create small size regular chunks and merge them into larger chunks subsequently either with regular or irregular shape. The larger chunks remain the I/O unit while the smaller tiles are the processing and data transfer unit. While the authors claim the regular-regular strategy to provide the best results, we question this conclusion when the implementation is optimized since the difference is negligible when compared to the elementary strategies. At a closer look, the two-level chunking strategies are a special case of the aligned chunking proposed in~\cite{furtado:tiling}. In the case of aligned chunking, the number of levels is dictated by the size of each chunk. Splitting and merging are executed recursively until chunks with the desired size are obtained. The cells assigned to the same chunk are stored in index-value format in no particular order, thus value-based operations require full chunk processing. The order in which chunks assigned to the same disk are linearized is not discussed at all. Chunk declustering across multiple disks is also treated superficially. The only strategies considered are based on data partitioning techniques from parallel databases. They include random assignment, round-robin, range, and block-cyclic as a special case of hash partitioning.

Significant attention is given to chunking strategies for overlapping operations---operations that have to access data from adjacent chunks in some corner cases. The strategies considered include post-processing, single-layer overlap, explicit run-time data request, and multiple-layer overlap. In post-processing, chunks are processed in parallel as usual. Only the corner cases require data to be moved around to finalize the processing. How often this happens and how much data have to be transferred are data-dependent. This approach requires no additional data storage and no special programming for the common case processing. With a good strategy for post-processing -- see GLADE~\cite{glade:demo} -- we argue that this approach is the most performant. Single and multiple-layer overlap require data to be replicated across multiple chunks. The extra data can be stored with the actual chunk or separately. The additional storage required is significant even when the overlap is minimal, e.g., 5\% on each side of a chunk. The advantage is that no data has to be transferred across chunks as long as the overlap contains such data. This is problematic since the overlap has to be determined ahead of processing. Explicit run-time data request involves moving data across chunks only when needed. Instead of moving the entire chunk, the two-level chunking strategy is used as an optimization to move only the required tiles. It is quite common in overlapping processing to create derived structures for the corner cases based on the raw data. Only these structures have to be merged together in the post-processing, thus an optimization is to transfer only the derived data. The authors do not consider this scenario in~\cite{arraystore}. They focus only on moving raw data around.

Based on the proposed chunking strategies, the workload operations are solved as follows. There is no support for value-based operations. All the cells inside the chunk have to be inspected for every filter operation, for example. Structural operations take advantage of the chunking. All the overlapping chunks are read from disk in the case of subsample. In the case of two-level strategies, only the overlapping tiles are processed. If an index is used for chunk storage, the chunks that have to be read can be identified faster---this actually depends on the number of chunks. The solution advocated for binary operations is to use the same chunking shape and to co-locate corresponding chunks at the same site. This is well-known to be optimal from parallel join algorithms. In the case of overlapping operations, the authors advocate for the two-level chunking strategy combined with explicit run-time data request.

The chunking strategies are incorporated into a storage manager that provides a standard pull-based interface centered around the \texttt{GetNext} method. We believe this is problematic for parallel processing since explicit requests have to be made by the clients. The only method to read multiple chunks in parallel is to resort to pre-fetching and local buffering in the storage manager. None of these issues are treated by the authors which seem to focus only on centralized processing---more evident from the experiments. Parallelism is obtained only by running multiple instances of the storage manager as separate processes, one for each node more likely. Given the complete lack of parallelism inside the storage manager, the applicability of the proposed solution for parallel array processing is questionable.

%%%%%%%%%%%%%%%%%%%%%%%%%%%%%%%%%%%%%%%%%%%%%%%%%%%%%%%%%%%%%%%%%%%%%%%%%%%%%%%%%%
%\input{state-of-the-art}
\section{State-of-the-Art in Array Processing}\label{sec:state-of-the-art}

The research surveyed in this work spans over the last two decades. The majority of the papers on array storage techniques, query languages, and array processing systems were published in the late 1990's. After a decline in interest in early 2000's, a renewed interest in array processing was triggered by the Extremely Large Databases (XLDB) series of workshops which initiated the development of the SciDB system targeted at large scale array processing. Thus, the recent work on array databases is mostly focused around the development of SciDB~\cite{scidb:demo}. In this section, we first examine how the initial research on array databases reflects in the novel SciDB system---we scrutinize the state of the SciDB eco-system at the time of writing this document. Then, we present some of the most recent ideas introduced in the context of array processing, including an in-depth discussion of in-situ data processing in the Map-Reduce~\cite{google:mapreduce} framework.

\subsection{SciDB}\label{ssec:sota:scidb}

It is quite fascinating to examine how more than a decade worth of research shapes the design of a novel prototype system. What exactly proved applicable enough to be preserved over time? And how much is "rediscovered" in the new iteration over the same topic? As expected, the simplified answer to these two questions is that, very likely, only intuitive ideas are preserved over time because they can be easily rediscovered. The reason for this might be that, at the end of the day, the most intuitive ideas are also the most practical ones---definitely, they are easier to implement. We exemplify these claims by analyzing what research ideas made it in the SciDB implementation at storage, processing, and query language level. And what new ideas are generated by the technological advancements and the new workloads.

\paragraph{Array storage.}
The SciDB storage system is partially modeled after ArrayStore~\cite{arraystore} (see Section~\ref{ssec:arraystore} for details). From all the alternative chunking strategies proposed in the literature, the most intuitive solution -- regular chunking with arbitrary chunk size -- is implemented. Chunk declustering across disks is implemented using standard parallel database data partitioning strategies rather than considering the more complicated declustering techniques presented in Section~\ref{ssec:array:storage:mapping-chunks}. SciDB is not the only system discarding more intricate storage solutions. ArrayDB~\cite{marathe:arrays} adopts the same strategy. The problem that receives considerable attention is the design of storage methods for overlap processing---this type of queries was not considered at all in the original array storage research. Multiple solutions are proposed that replicate data across chunks. Array decomposition and chunk compression (see Section~\ref{ssec:scidb} for details) are new problems orthogonal to standard array storage introduced by the SciDB design. The proposed solutions are extensions from columnar databases rather than novel research results.

\paragraph{Array processing.}
The original SciDB design~\cite{scidb:demo} recognized the diversity in array manipulations and modeled the entire processing as a sequence of UDF invocations. While this approach might work well in a centralized environment, the lack of semantics in the UDF definition -- any user code can be part of a UDF -- creates problems in scheduling the execution over a shared-nothing parallel architecture. Optimizing a given execution plan is also problematic without further information on UDF properties since it is impossible to change the execution order. As a result, the SciDB implementation focuses on parallelizing each UDF independently. This is done only across processes though, with little attention paid multi-threading parallelism. Except these, array processing in SciDB is quite standard.

\paragraph{Array query languages.}
As already emphasized in Section~\ref{sec:query-lang}, there is no commonly accepted array query language to date. Consequently, an important topic in the SciDB design is to define an array algebra and query language. While there are multiple efforts initiated by the XLDB community, there is little consensus beyond the AFL and AQL languages introduced in Section~\ref{ssec:query-lang:SciDB}.

In~\cite{aql:algebra}, a proposal for an array algebra in the SciDB context is made. Arrays are formalized as 3-tuples of the form (Box, Valid, Content), where Box represents the domain of the array with fixed bounds on all dimensions, Valid is a boolean map indicating which cells have valid values, and Content is a function providing the values for the array cells. This is the first algebra that represents cell validity explicitly. The benefit is that both dense and sparse arrays can be formulated within the same algebra constructs. Moreover, having a fixed-size domain simplifies implementation considerably. Based on this array formalization, an entire suite of operators is defined. Given the representation of an array as a 3-tuple, a new array is created by each operator, with a corresponding new 3-tuple. Operators define mappings between the original 3-tuple components and the new components. Without going into details, the following operators are defined in the algebra: \textit{Rename} dimension or attribute; \textit{Shift} origin; \textit{Rebox} changes the dimension sizes; \textit{Filter} invalidates some array cells based on a content-only predicate; \textit{Fill} transforms all the invalid cells to valid and assigns them a value based on a function provided as argument; \textit{Apply} applies a function passed as argument to each valid cell of an array; \textit{Combine} combines the content of two arrays having the same shape, but not necessarily the same validity; \textit{InnerDJoin} and \textit{InnerEJoin} are join operators that work on dimensions, and dimensions and attributes, respectively; \textit{Reduce} generates a reduced version of an array by aggregating over one or more dimensions passed as arguments. These operators work on each of the three components in the array formalization---dimensions, content, and validity. Or on combinations of them. As the authors point out, not every array operation can be expressed with this algebra, e.g., promote attributes into dimensions, or demote a dimension into an attribute. What is not emphasized enough though, is the expressive power of the algebra or the completeness. Given this is only a draft, these shortcomings are understandable. Compared to previously proposed array algebras~\cite{AQL,rasdaman,marathe:arrays}, the SciDB proposal has both similarities and differences. It specifies operations on dimensions and cell content. In addition, it also contains operations on validity and joins between arrays. Similar to the other algebras, generality and extensibility are supported through second-order operators taking user-defined functions (UDF) as arguments. Different from other algebras, the dimensionality of the arrays passed as arguments to operators is strictly enforced. The implementation and optimization of this algebra is an interesting exercise to pursue.

ArrayQL~\cite{aql:syntax} is defined as an array creation and query language based on the algebra proposed in~\cite{aql:algebra} and presented above. It is highly reminiscent of SQL and contains only two statements---CREATE ARRAY to create arrays at the schema level and SELECT FROM to query arrays. There are no statements to insert/update data in arrays. ArrayQL queries take as input only arrays. The output can be either a new array -- with dimensions specified explicitly in the query as brackets -- or a relation---without any ordering constraint. Ranges on dimensions can be specified both for the input and the output arrays. In the case of input arrays, ranges correspond to sub-arrays, while in the case of the result array, ranges implement the \textit{Shift} operator. If no ranges are provided, the entire dimension ranges of the input array(s) are automatically inherited. Structural joins between two arrays are specified by enumerating the arrays in the FROM clause and matching the dimensions' names. Overall, algebra operators are mostly implemented through index mappings. Not all the proposed operators are specified in the language though. And not all the operations possible in the language by the means of intricate index mappings are part of the array algebra~\cite{aql:algebra}. There is considerable work to be done to bring these two proposals~\cite{aql:algebra,aql:syntax} at the level of complete array query language specifications.

\subsection{Array Processing in Map-Reduce}\label{ssec:hadoop}

Map-Reduce~\cite{google:mapreduce} is a framework for simplified parallel processing over massive datasets. It has two defining characteristics. It is scalable up to thousands of computing nodes. And it provides fault tolerance in the face of node failures. Moreover, Hadoop~\cite{hadoop} -- the open source Map-Reduce implementation -- supports the execution of customizable user tasks over arbitrary formatted datasets. All these properties make Hadoop Map-Reduce an attractive platform for large scale scientific data processing -- especially in-situ data processing -- even though its efficiency is known to be abysmally poor when compared to parallel databases.

In this section, we take a closer look at how large scale scientific processing over massive datasets is implemented in Hadoop. Specifically, we discuss in details multiple systems that extend Hadoop with support for arrays stored in well-known scientific file formats, e.g., FITS, HDF5, (Net)CDF. The efficiency mismatch is not that important in this scenario since databases do not support arrays natively nor can they operate over in-situ data without significant performance degradation.

\paragraph{HAMA.}
HAMA~\cite{hama} is a framework for matrix operations implemented on top of Hadoop Map-Reduce. It supports various matrix operations, including matrix multiplication and the conjugate gradient method for solving linear equation systems. Matrices are stored inside the HBase database extension to HDFS -- the distributed file system under Hadoop -- while operations are implemented as standard Map-Reduce programs. Thus, there is nothing special for array processing in HAMA---both at storage and execution level, respectively.

\paragraph{SciHadoop \& SIDR.}
SciHadoop~\cite{scihadoop} introduces multiple partitioning strategies for array data over HDFS chunks in order to optimize the computation of holistic aggregates such as the median. Alternative partitioning strategies to the default chunking mechanism are required because the array structure has to be considered for optimal execution. Whenever possible, the array is partitioned such that the holistic function can be computed at a single node with a holistic combiner. It metadata are available, HDFS chunks are created to align perfectly with the logical array partitions. As a result, the processing of each partition is restricted to the corresponding chunk. When metadata are not available, a bottom-up partitioning strategy is employed in which fine-grained sub-arrays are first created and then they are merged together into chunks such that the number of partitions crossing chunk boundaries is minimized. The last optimization targets the scheduler. It is a simple indexing strategy that enforces the processing of the chunks required by the query. By default, Hadoop reads all the chunks of the array and only then applies the range predicates in the Map function. All these optimizations are implemented for the NetCDF file format and then evaluated against the default Hadoop chunking scheme.

SIDR~\cite{sidr} introduces additional optimizations to SciHadoop, similar in spirit to the array processing mechanism implemented in the Titan system~\cite{titan}. The main problem with the Map-Reduce implementation in Hadoop is that a Reducer can start processing only after all the Mappers finish their execution. This is due to the assumption that all the Mappers contribute data for all the Reducers and is supported by the hash-based shuffling of the intermediate data. This assumption does not hold for array processing -- grid image manipulation in particular -- where an output cell typically depends only on the neighboring cells. The main contribution made by SIDR is to extend SciHadoop with functionality that allows the Reducers to determine when the Mappers contributing data for their result did finish processing, thus to start processing much earlier. The Reducers start working gradually, as the corresponding Mappers finish, rather than all at the same time -- when all the Mappers finish. A secondary effect of this strategy is the reduction in the overall execution time due to reduced network traffic.

\paragraph{SciMATE.}
SciMATE~\cite{scimate} is a framework for mapping any scientific file format into the chunking format supported by MATE---a system for scientific data processing with a similar interface to Hadoop Map-Reduce. An API consisting in a series of functions that convert data in the original format to the MATE format are introduced together with multiple optimizations to enhance the conversion process and to speed-up the execution. The main problem is what functions for the original format to invoke and what parameters to use. Since the answer is highly-dependent on the queries and on the specific file format, the solution proposed emphasizes flexibility and extensibility.

\subsection{Novel Research Directions}\label{ssec:sota:new-dir}

SciDB~\cite{scidb:demo} is a scientific data management system optimized for array processing. Modern scientific processing encompasses considerably more than data storage and query execution, as it is clearly emphasized by the SciDB design objectives~\cite{scidb:ssdbm-11}. In the following, we present the most recent steps taken in order to achieve these goals. They address in-situ data processing, versioning, provenance, and benchmarking. While the first three are required features for a modern scientific data processing, benchmarking in the context of array databases is a topic rarely addressed. Essentially, there is no benchmark for array processing systems.

\subsubsection{In-situ data processing}\label{sssec:in-situ}

Scientific data is generated in a variety of file formats, including FITS, HDF5, CDF, and many others. In order to process these data inside a database, data has to be first loaded inside the system. When the size of the data is large, the loading process takes significant amounts of time since it requires reading all the data from disk in the original format, mapping the data into the database internal representation, and writing the transformed data back to disk. Moreover, the space taken by data on disk doubles since the same data is stored in two different formats---this is prohibitively large in the case of terabytes to petabytes of data. In-situ data processing eliminates the loading phase completely and allows the database to operate on the original file format while still preserving all the upstream database functionality. To make this possible, a data converter mapping data in the original format to the database representation has to be integrated at the disk access layer. The converter guarantees that, once they touch the database, data are in the internal representation. The typical functionality of the converter is to only generate the data. The entire processing logic is kept in the database. The converter is also stateless, regenerating data from scratch for every query, without "remembering" any of the previous work. An alternative to this solution is to combine conversion and data loading. When data are first accessed in the original format, they are converted in the internal representation, passed to the query evaluation plan, and stored in the internal representation. Subsequent queries accessing the already loaded data read them from the database rather than from the original format. Thus, conversion is executed only the first time data are accessed, not every time. Only the first query experiences a performance hit, not all the queries.

\paragraph{NoDB.}
NoDB~\cite{nodb} is a modern converter for in-situ data processing with enhanced functionality. It maps only the required data for the query at hand. It implements logic for database operators at the storage level. And it has "memory", thus it is capable of reusing shared work across similar repeated queries. The objective of NoDB is to minimize the access time to raw data over time without loading it into the database. It accomplishes this by combining multiple techniques. Mapping the original file format into the internal database representation requires parsing and tokenizing the original file. If the file is in text format, e.g., CSV -- the work case in NoDB -- tuple and attribute delimiters have to be identified. The text representation of each attribute is then transformed into its corresponding binary representation. This has to be done for each tuple and each attribute in the general loading case.

In NoDB, both the parsing and the tokenization are selective and query-driven. This means that only the attributes required in the query at hand are parsed and tokenized. In other words, the projection operator is pushed down from the database into the converter. Moreover, selections are also pushed down into the NoDB converter. They are evaluated as soon as the corresponding attributes are tokenized. If the selection condition is not satisfied, the tuple is immediately discarded, without further parsing and tokenization, and without passing it at all into the database engine. To further reduce access time to raw data, a positional map recording the starting positions of all the parsed attributes is built on-the-fly, during query processing. This is equivalent to an index on the raw data and it is used to identify directly the positions of the attributes in the map without parsing the entire tuple. In the extreme case, the map contains an entry for each tuple and each attribute. In the general case though, the map contains entries only for the queried attributes. During query execution, the positions of all the accessed attributes are first computed and only then they are fetched and tokenized from the raw file. This clean ordering of operations, although increases the memory requirement, it speeds-up processing by providing a better mix of instruction to the CPU. The positional map is adaptive and maintained based on the queries that are executed. In addition to the positional map index, NoDB also caches partial or full tuples in memory. If the cached tuples are required by a subsequent query, they are retrieved from the cache in binary format. No more parsing and tokenization have to be executed. The cache is used on top of the positional map index. To push things further, even statistics computation is implemented in the NoDB converter and then fed into the database for enhancing query optimization.

When all these techniques are combined together, in-situ data processing becomes almost as efficient as executing queries inside the database on previously loaded data, without actually executing the loading. The performance for the first few queries is not that great -- the positional map and the cache have to be populated in this phase -- but subsequent queries are executed even faster than on the internal database representation in some cases.

\paragraph{Data Vaults.}
Data Vaults~\cite{data-vaults} is a scientific data processing system that provides a symbiosis between a DBMS and external file repositories. Data are stored in the original format in a file repository---they are not loaded into the database apriori. Metadata and pointers to data -- paths to files -- are stored in relational tables. To allow data processing inside the database, raw data are mapped into internal array structures -- raw data are arrays -- and then processed through SciQL~\cite{SciQL} query statements. To make this possible, wrappers from the original file format to the internal representation have to be implemented inside the database. In order not to execute the conversion for every query, just-in-time loading and caching are performed to preserve once converted data inside the database.

Data Vaults are implemented on top of the MonetDB~\cite{monetdb:overview} SciQL~\cite{SciQL} stack. Three components are added to the system. The wrapper manager is in charge of the conversion between external file formats and the internal MonetDB representation. There is a wrapper for each different scientific file format. The wrappers act like data access methods with possibility for just-in-time loading. The cache manager is in charge of maintaining the data loaded into the database consistent with the raw data. And the optimizer decides what data to execute queries upon---cached data -- when available -- or raw data with required conversion through wrappers. All these ideas are conceptual. There are little implementation details and no real experimental results to validate the performance of the approach.

\subsubsection{Versioning}\label{sssec:version}

The main idea behind the versioning concept is the requirement imposed by scientists to never modify data in place. Every modification has to generate another version of the original data. Some of the versions are given names, while the majority are identified based on a sequential identifier assigned automatically by the system whenever a modification takes place. Maintaining all the states data are transformed through allows novel types of queries, i.e., time travel queries. These are queries that retrieve a particular version of the data at a given instance in time or in a given named version and queries that return all the states (a subset of) the data go through across all or a subset of the versions. Abstractly, array versioning corresponds to adding a new time dimension to the original array, while time travel queries are either slice or range queries along the time dimension.

There are multiple research problems that have to be addressed by an array versioning system. The first, and most important, is how to minimize the storage space occupied by all the versions of the same array. The naive solution is to materialize each version independently of other versions. This results in storage proportional with the number of versions and continuously increasing as new versions are created. The observation allowing for alternative solutions is that new versions typically change only a small portion of the array, while maintaining the majority of the array unchanged. Thus, storing only the modifications has the potential to save a significant amount of disk space. Alternatively, a delta array that contains the difference between a base array and a version at each cell can be generated. Under the assumption that the differences are small, thus require fewer bytes to encode, storage can be saved. The next question is which version to materialize? Under the assumption that the newest version is queried more frequently, the newest version is always materialized. To reduce the number of previous versions that have to be re-encoded based on each new version, older versions are maintained as the difference or delta from the immediately successive version. A chain of version deltas results in which version one is materialized as the difference from version two, version two as the difference from version three, and so on, with the second newest version encoded as the delta from the newest version which is fully materialized. When versions are maintained as deltas, answering time travel queries is more complicated since computation -- combine delta(s) with a materialized version -- is required to get the result. Efficiently answering time travel queries with deltas is thus another important research problem in the context of versioning.

There are two solutions proposed for building, maintaining, and querying versioned arrays in the literature~\cite{arrversion,timearr}. They both propose storage managers for versioned arrays in the context of SciDB~\cite{scidb:demo}. As such, they build on top of the SciDB array storage architecture which decomposes a multi-valued array into multiple arrays with a single value in each cell. Each such array is further chunked into regular chunks that are round-robin partitioned across the available disks. Thus,~\cite{arrversion,timearr} consider the simplified problem of versioning single-valued arrays regularly chunked. The main idea is to store a single materialized version of the chunk and all its deltas inside the same chunk file without considering alternative chunking strategies as new versions are created.

The problem considered in~\cite{arrversion} is how to optimally encode a series of $n$ consecutive versions given at once rather than created incrementally under the assumption that queries across all the versions are equally probable---materializing the last version is not the optimal solution in this case. The goal is to determine which versions to materialize and based on which materialized versions to create deltas for the non-materialized versions. To find the exact solution for this problem requires considering all possible $n^{n}$ materialization combinations which is impractical even for small values of $n$. The proposed algorithm finds optimal encodings without exploring all the materialization options. Only a single materialization is considered for each version---no replicas are allowed. The algorithm starts by computing a materialization matrix with the cost of materializing each version on the diagonal and the cost of encoding version $i$ as a delta of version $j$ at position $\left(i, j\right)$. The matrix is symmetric and is computed using only a random sample of the array cells, not the entire arrays. This matrix is subsequently treated as the adjacency matrix of a complete directed graph in which nodes are the versions and edges have the following significance. An edge between node $i$ and $j$ corresponds to storing version $i$ as a delta of version $j$. An edge from $i$ to $i$ corresponds to materializing version $i$. Edges have the materialization costs attached. Finding the optimal materialization strategy corresponds to identifying the minimum spanning forest in the graph possibly with more than one version materialized. When the cost of each edge $\left(i, j\right)$ is smaller than the materialization cost for all the versions, the minimum spanning forest is actually the minimum spanning tree of the undirected graph corresponding to the complete graph built with the materialization matrix. Computing the minimum spanning tree is a problem with well-known solutions. The optimal materialization can be immediately determined by selecting the version with the lowest materialization cost as the only materialized version and then computing deltas by traversing the tree starting at the root---the materialized version. When there is a version for which the materialization cost is smaller than some delta cost, the minimum spanning tree identified previously is considered for splitting into a forest if certain conditions are met. There is no proof in the paper that such a split still generates the optimal materialization.

In~\cite{timearr}, the authors consider multiple problems in the context of a backward delta versioning system specialized for arrays. In the scenario considered, versions of the same array are created incrementally, one after another. The assumption is that the most recent version is queried considerably more often than the others. Moreover, the probability of querying a version decreases with its age. Thus, in backward delta versioning, the most recent version is always fully materialized. Each other version is stored as a delta difference from the immediately successive version. When a new version is created, only the second newest version has to be delta encoded from the newly created version. The query execution time increases with the age of the queried version since a larger number of deltas have to be computed. In order to minimize the disk space allocated to storing the entire set of array versions, multiple approaches are combined. Delta encoding is applied at tile level, where tiles are a second level of regular chunking inside regular chunks---these are the same authors of the ArrayStore~\cite{arraystore} paper. Different tiles in the same chunk can use different length delta encodings. Compressed bit-masks encoding only the tiles that change from one version to another are stored in the chunk metadata. And in the tile metadata for the cells, respectively. The bit-masks are run-length encoded to reduce the storage space they use. To address the problem of increasing query response time with the version age, skip links, i.e., tile-level delta encodings between non-consecutive similar versions, are built lazily while evaluating queries. This results in much faster response time if a similar query has to be answered. A skip link always has smaller disk footprint than the regular consecutive delta encoding---the two tiles have more similar content resulting in better compression. Only skip links to the most recent version are considered, only when querying an old version, and only between versions newer than the queried version and the newest version. Skip links are also used to provide faster approximate answers to other non-similar queries. Approximation corresponds to answering a query using a different -- preferably far in time and similar in content -- version than the one requested. It is controlled by a maximum acceptable error, specified by the user and computed as an aggregate value at chunk or tile level. Instead of following the delta path all the way to the queried version, delta reconstruction can be stopped as early as the tolerable error is satisfied. This requires the ability to compute the approximation error between any two versions. In the solution proposed, the error is pre-computed at version creation for a set of measure functions only between consecutive versions and between the newly inserted version and the oldest version. The functions are computed at the tile and chunk level---the error measure has to be distributive to allow direct computation at chunk level from the corresponding tile values.

\subsubsection{Provenance}\label{sssec:provenance}

It is typical in scientific processing to have workflows of operators that take input arrays, apply multiple transformations, and generate an output array. Given an output cell, it is quite common to ask what are the cells in the input array on which the output cell depends. Or the inverse equivalent query, what are the output cells which depend on a given input cell. To complicate the problem further, these types of queries can be asked for any pair of operators in the workflow, not necessarily the source(s) and the result(s). To answer these queries after the workflow is processed without entirely re-executing it, lineage data have to be stored for each operator in the workflow, in both directions. If such data are generated at cell level for all the arrays in the workflow, it is very likely that the amount of additional space might overpass the original data multiple times. And it is not guaranteed that answering the provenance queries is going to be faster than re-executing the workflow. Determining which data to materialize and which to recompute is the fundamental question in array provenance.

\paragraph{SubZero.}
SubZero~\cite{subzero} is a prototype system for managing array provenance data. It is based on the idea of region lineage as an intermediate level to generate and store lineage data based on locality. SubZero introduces a lineage API that allows developers to expose lineage data from user-defined functions and operators through mapping functions. Given a workflow consisting of a series of operators, SubZero uses an optimization framework to select the optimal strategy to generate lineage data for a given workload. Multiple strategies to generate lineage data are considered for each operator and their corresponding cost. SubZero can record and store the lineage data at workflow runtime or it can decide that it is more efficient to re-execute the workflow, case in which the lineage data are generated only during the execution of the provenance query---after answering the query, the provenance data are discarded.

Each operator in the workflow can support multiple types of lineage data. Black-box lineage data record only the input and output arrays of each operator together with the execution parameters. Cell-level lineage records pairs of (input, output) array cells, where the output cell is dependent on the input cell. An input/output cell can be part of many pairs. Region lineage is similar to cell-level at a coarser granularity---all-to-all cell-level lineage applies between every cell in the input region and every cell in the output region. Multiple strategies to generate and store the region lineage data are presented. Operators can implement one or more strategies. In full lineage, all the region pairs are stored explicitly. In mapping lineage, only two mapping functions -- forward and backward -- have to be specified for an operator. Each function specifies the output coordinates as a function of the input cell coordinates. The functions do not depend on cell content. They are structural array operators. No lineage data are stored in this case. At query time, the lineage can be computed for each cell based on the coordinates. Two more strategies derived from mapping lineage are also introduced in~\cite{subzero}. We do not discuss them in this document.

\subsubsection{Benchmarking}\label{sssec:benchmark}

\paragraph{SS-DB.}
The \textit{Standard Science DBMS Benchmark} (SS-DB)~\cite{ssdb} is a general benchmark to evaluate the performance of array processing systems. It is based on processing astronomical images and inherits many features from the Sloan Digital Sky Survey project~\cite{sloan-sky-survey}. Unlike Sloan Digital Sky Survey which operates on relational data obtained as a bi-product of astronomical observations, SS-DB is more general and contains a full spectrum of operations ranging from raw data processing to the creation and querying of derived observation data. These operations manipulate array-oriented data through relatively sophisticated user-defined functions. SS-DB contains queries on 1-D arrays (e.g., polygon boundaries), dense/sparse 2-D arrays (e.g., imagery data/"cooked" data), and 3-D arrays (e.g., trajectories in space and time). Data loading is also part of the benchmark specification. The benchmark queries involve rather intricate operations that go beyond standard operators on arrays. The only solution to implement the queries is to write custom code that can be integrated with a dedicated storage system, i.e., user-defined functions (UDF)---the SciDB implementation of the benchmark consists of a series of UDFs for the SS-DB queries and the cooking operations. As a result, it is questionable if SS-DB measures the performance of a set of well-defined array primitives or only the implementation quality of a set of UDFs. Moreover, the experimental results presented in~\cite{ssdb} which compare SciDB against MySQL focus on general features characteristic to a modern parallel storage and processing engine such as columnar storage, compression, and multi-threaded parallelism rather than on specific array operations. Since MySQL lacks all these features, the two orders of magnitude difference in performance in favor of SciDB is expected.

%%%%%%%%%%%%%%%%%%%%%%%%%%%%%%%%%%%%%%%%%%%%%%%%%%%%%%%%%%%%%%%%%%%%%%%%%%%%%%%%%%
%\input{conclusions}
\section{Conclusions}\label{sec:conclusions}

There is a renewed interest in array data processing in the context of Big Data since scientific investigation is one of the most important generators of massive amounts of data. There is also a lack of a unified resource that summarizes and analyzes array processing research over its long existence. In this survey, we provide this missing resource as a guide for current and new research in array processing. We present the problem from a database perspective, thus we focus our attention on clarifying the subtle differences between the relational data model and ordered arrays. The survey is organized along three main topics. Array storage discusses all the aspects related to how to partition arrays into chunks for storage across a single or multiple disks. Chunk size, organization, and ordering on disk are some of the most important problems we address. The identification of a reduced set of array operators to form the foundation for an array query language is analyzed across multiple proposals. None of them is unanimously accepted yet. And last, but not least, we investigate how are the research ideas implemented in real systems for array processing. We conclude the survey by presenting the most recent and future research topics in array data processing.

%%%%%%%%%%%%%%%%%%%%%%%%%%%%%%%%%%%%%%%%%%%%%%%%%%%%%%%%%%%%%%%%%%%%%%%%%%%%%%%%%%
\bibliographystyle{abbrv}

\begin{thebibliography}{10}

\bibitem{postgresql}
{PostgreSQL}.
\newblock \url{http://www.postgresql.org/}.
\newblock [Online; accessed July 2012].

\bibitem{nodb}
I.~Alagiannis, R.~Borovica, M.~Branco, S.~Idreos, and A.~Ailamaki.
\newblock {NoDB: Efficient Query Execution on Raw Data Files}.
\newblock In {\em {Proceedings of 2012 ACM SIGMOD International Conference on
  Management of Data}}, pages 241--252, 2012.

\bibitem{baumann:vldbj}
P.~Baumann.
\newblock {On the Management of Multi-Dimensional Discrete Data}.
\newblock {\em VLDB Journal (VLDBJ)}, 4(3):401--444, 1994.

\bibitem{baumann:ngits}
P.~Baumann.
\newblock {A Database Array Algebra for Spatio-Temporal Data and Beyond}.
\newblock In {\em Proceedings of 1999 NGITS International Workshop on Next
  Generation Information Technologies and Systems}, pages 76--93, 1999.

\bibitem{rasdaman}
P.~Baumann, A.~Dehmel, P.~Furtado, R.~Ritsch, and N.~Widmann.
\newblock {The Multidimensional Database System RasDaMan}.
\newblock In {\em {Proceedings of 1998 ACM SIGMOD International Conference on
  Management of Data}}, pages 575--577, 1998.

\bibitem{baumann:compare-array-algebra}
P.~Baumann and S.~Holsten.
\newblock {A Comparative Analysis of Array Models for Databases}.
\newblock In {\em Proceedings of 2011 FGIT-DTA/BSBT}, pages 80--89, 2011.

\bibitem{monetdbX100}
P.~Boncz, M.~Zukowski, and N.~Nes.
\newblock {MonetDB/X100: Hyper-Pipelining Query Execution}.
\newblock In {\em Proceedings of 2005 CIDR Conference on Innovative Database
  Research}, pages 225--237, 2005.

\bibitem{sidr}
J.~Buck, N.~Watkins, G.~Levin, A.~Crume, K.~Ioannidou, S.~Brandt, C.~Maltzahn,
  and N.~Polyzotis.
\newblock {SIDR: Efficient Structure-Aware Intelligent Data Routing in
  SciHadoop}.
\newblock Technical Report UCSC-TR-SOE-12-08, UC Santa Cruz, Jack Baskin School
  of Engineering, 2012.

\bibitem{scihadoop}
J.~B. Buck, N.~Watkins, J.~LeFevre, K.~Ioannidou, C.~Maltzahn, N.~Polyzotis,
  and S.~Brandt.
\newblock {SciHadoop: Array-based Query Processing in Hadoop}.
\newblock In {\em {Proceedings of 2011 SC International Conference for High
  Performance Computing, Networking, Storage and Analysis}}, pages 66:1--66:11,
  2011.

\bibitem{t2}
C.~Chang, A.~Acharya, A.~Sussman, and J.~H. Saltz.
\newblock {T2: A Customizable Parallel Database for Multi-Dimensional Data}.
\newblock {\em SIGMOD Rec.}, 27(1):58--66, 1998.

\bibitem{titan}
C.~Chang, B.~Moon, A.~Acharya, C.~Shock, A.~Sussman, and J.~H. Saltz.
\newblock {Titan: A High-Performance Remote Sensing Database}.
\newblock In {\em {Proceedings of 1997 IEEE ICDE International Conference on
  Data Engineering}}, pages 375--384, 1997.

\bibitem{glade:demo}
Y.~Cheng, C.~Qin, and F.~Rusu.
\newblock {GLADE: Big Data Analytics Made Easy}.
\newblock In {\em Proceedings of 2012 ACM SIGMOD International Conference on
  Management of Data}, pages 697--700, 2012.

\bibitem{sram}
R.~Cornacchia, S.~H{\'e}man, M.~Zukowski, A.~P. de~Vries, and P.~Boncz.
\newblock {Flexible and Efficient IR using Array Databases}.
\newblock {\em VLDB Journal (VLDBJ)}, 17:151--168, 2008.

\bibitem{ssdb}
P.~Cudre-Mauroux, H.~Kimura, K.-T. Lim, J.~Rogers, S.~Madden, M.~Stonebraker,
  S.~B. Zdonik, and P.~G. Brown.
\newblock {SS-DB: A Standard Science DBMS Benchmark}.
\newblock \url{http://www.xldb.org/science-benchmark/}.
\newblock [Online; accessed August 2012].

\bibitem{scidb:demo}
P.~Cudre-Mauroux, H.~Kimura, K.-T. Lim, J.~Rogers, R.~Simakov, E.~Soroush,
  P.~Velikhov, D.~L. Wang, M.~Balazinska, J.~Becla, D.~DeWitt, B.~Heath,
  D.~Maier, S.~Madden, J.~Patel, M.~Stonebraker, and S.~Zdonik.
\newblock {A Demonstration of SciDB: A Science-Oriented DBMS}.
\newblock {\em PVLDB}, 2(2):1534--1537, 2009.

\bibitem{google:mapreduce}
J.~Dean and S.~Ghemawat.
\newblock {MapReduce: Simplified Data Processing on Large Clusters}.
\newblock {\em Commun. ACM}, 51(1):107--113, 2008.

\bibitem{faloutsos:hcam}
C.~Faloutsos and P.~Bhagwat.
\newblock {Declustering Using Fractals}.
\newblock In {\em {Proceedings of 1993 International Conference on Parallel and
  Distributed Information Systems}}, pages 18--25, 1993.

\bibitem{furtado:tiling}
P.~Furtado and P.~Baumann.
\newblock {Storage of Multidimensional Arrays Based on Arbitrary Tiling}.
\newblock In {\em {Proceedings of 1999 IEEE ICDE International Conference on
  Data Engineering}}, pages 480--489, 1999.

\bibitem{sparse:bitmap-index}
S.~Goil and A.~N. Choudhary.
\newblock {Sparse Data Storage Schemes for Multidimensional Data for OLAP and
  Data Mining}.
\newblock Technical Report CPDC-TR-9801-005, Center for Parallel and
  Distributed Computing, Northwestern University, 1997.

\bibitem{hadoop}
Hadoop.
\newblock \url{http://hadoop.apache.org/}.
\newblock [Online; accessed July 2011].

\bibitem{howe:array-algebra}
B.~Howe and D.~Maier.
\newblock {Algebraic Manipulation of Scientific Datasets}.
\newblock In {\em {Proceedings of 2004 VLDB International Conference on Very
  Large Databases}}, pages 924--935, 2004.

\bibitem{monetdb:overview}
S.~Idreos, F.~Groffen, N.~Nes, S.~Manegold, K.~S. Mullender, and M.~L. Kersten.
\newblock {MonetDB: Two Decades of Research in Column-Oriented Database
  Architectures}.
\newblock {\em IEEE Data Eng. Bull.}, 35(1):40--45, 2012.

\bibitem{data-vaults}
M.~Ivanova, M.~L. Kersten, and S.~Manegold.
\newblock {Data Vaults: A Symbiosis between Database Technology and Scientific
  File Repositories}.
\newblock In {\em Proceedings of 2012 SSDBM International Conference on
  Scientific and Statistical Database Management}, pages 485--494, 2012.

\bibitem{jagadish:hilbert-clustering}
H.~Jagadish.
\newblock {Linear Clustering of Objects with Multiple Attributes}.
\newblock In {\em {Proceedings of 1990 ACM SIGMOD International Conference on
  Management of Data}}, pages 332--342, 1990.

\bibitem{SciQL-edbt}
M.~L. Kersten, Y.~Zhang, M.~Ivanova, and N.~Nes.
\newblock {SciQL, A Query Language for Science Applications}.
\newblock In {\em Proceedings of 2011 AD EDBT/ICDT Array Databases Workshop},
  pages 1--12, 2011.

\bibitem{aquery}
A.~Lerner and D.~Shasha.
\newblock {AQuery: Query Language for Ordered Data, Optimization Techniques,
  and Experiments}.
\newblock In {\em {Proceedings of 2003 VLDB International Conference on Very
  Large Databases}}, pages 345--356, 2003.

\bibitem{AQL}
L.~Libkin, R.~Machlin, and L.~Wong.
\newblock {A Query Language for Multidimensional Arrays: Design,
  Implementation, and Optimization Techniques}.
\newblock In {\em Proceedings of 1996 ACM SIGMOD International Conference on
  Management of Data}, pages 228--239, 1996.

\bibitem{aql:syntax}
K.-T. Lim, D.~Maier, J.~Becla, M.~Kersten, Y.~Zhang, and M.~Stonebraker.
\newblock {Array QL Syntax}.
\newblock
  \url{http://www.xldb.org/wp-content/uploads/2012/09/ArrayQL-Draft-4.pdf}.
\newblock [Online; accessed January 2013].

\bibitem{declustering:graph}
D.-R. Liu and S.~Shekhar.
\newblock {A Similarity Graph-based Approach to Declustering Problems and Its
  Application towards Parallelizing Grid Files}.
\newblock In {\em {Proceedings of 1995 IEEE ICDE International Conference on
  Data Engineering}}, pages 373--381, 1995.

\bibitem{aql:algebra}
D.~Maier.
\newblock {ArrayQL Algebra: version 3}.
\newblock
  \url{http://www.xldb.org/wp-content/uploads/2012/09/ArrayQL_Algebra_v3+.pdf}.
\newblock [Online; accessed January 2013].

\bibitem{maier:order}
D.~Maier and B.~Vance.
\newblock {A Call to Order}.
\newblock In {\em Proceedings of 1993 PODS Symposium on Principles of Database
  Systems}, pages 1--16, 1993.

\bibitem{marathe:arrays}
A.~P. Marathe and K.~Salem.
\newblock {Query Processing Techniques for Arrays}.
\newblock {\em VLDB Journal (VLDBJ)}, 11(1):68--91, 2002.

\bibitem{mckinsey}
{McKinsey Global Institute}.
\newblock {Big Data: The Next Frontier for Innovation, Competition, and
  Productivity}.
\newblock
  \url{http://www.mckinsey.com/insights/mgi/research/technology_and_innovation%
/big_data_the_next_frontier_for_innovation}, May 2011.
\newblock [Online; accessed August 2012].

\bibitem{moon:minimax}
B.~Moon, A.~Acharya, and J.~Saltz.
\newblock {Study of Scalable Declustering Algorithms for Parallel Grid Files}.
\newblock In {\em Proceedings of 1996 Parallel Processing Symposium}, pages
  434--440, 1996.

\bibitem{cart-prod-file:declustering}
B.~Moon and J.~H. Saltz.
\newblock {Scalability Analysis of Declustering Methods for Multidimensional
  Range Queries}.
\newblock {\em Transactions on Knowledge and Data Engineering (TKDE)},
  10(2):310--327, 1998.

\bibitem{white-house}
{Office of Science and Technology Policy, Executive Office of the President}.
\newblock {Obama Administration Unveils Big Data Initiative}.
\newblock
  \url{http://www.whitehouse.gov/sites/default/files/microsites/ostp/big_data_%
press_release_final_2.pdf}, March 29, 2012.
\newblock [Online; accessed August 2012].

\bibitem{rotem:optimal-chunking}
E.~J. Otoo, D.~Rotem, and S.~Seshadri.
\newblock {Optimal Chunking of Large Multidimensional Arrays for Data
  Warehousing}.
\newblock In {\em {Proceedings of 2007 ACM DOLAP International Workshop on Data
  Warehousing and OLAP}}, pages 25--32, 2007.

\bibitem{divy:cyclic-declustering-2D}
S.~Prabhakar, K.~Abdel-Ghaffar, D.~Agrawal, and A.~E. Abbadi.
\newblock {Cyclic Allocation of Two-Dimensional Data}.
\newblock In {\em {Proceedings of 1998 IEEE ICDE International Conference on
  Data Engineering}}, pages 94--101, 1998.

\bibitem{glade:osr}
F.~Rusu and A.~Dobra.
\newblock {GLADE: A Scalable Framework for Efficient Analytics}.
\newblock {\em OS Review}, 46(1), 2012.

\bibitem{sarawagi:chunking}
S.~Sarawagi and M.~Stonebraker.
\newblock {Efficient Organization of Large Multidimensional Arrays}.
\newblock In {\em {Proceedings of 1994 IEEE ICDE International Conference on
  Data Engineering}}, pages 328--336, 1994.

\bibitem{arrversion}
A.~Seering, P.~Cudre-Mauroux, S.~Madden, and M.~Stonebraker.
\newblock {Efficient Versioning for Scientific Array Databases}.
\newblock In {\em {Proceedings of 2012 IEEE ICDE International Conference on
  Data Engineering}}, pages 1013--1024, 2012.

\bibitem{hama}
S.~Seo, E.~J. Yoon, J.~Kim, S.~Jin, J.-S. Kim, and S.~Maeng.
\newblock {HAMA: An Efficient Matrix Computation with the MapReduce Framework}.
\newblock In {\em {Proceedings of 2010 IEEE CLOUDCOM International Conference
  on Cloud Computing Technology and Science}}, pages 721--726, 2010.

\bibitem{timearr}
E.~Soroush and M.~Balazinska.
\newblock {Time Travel in a Scientific Array Database}.
\newblock In {\em {Proceedings of 2013 IEEE ICDE International Conference on
  Data Engineering}}, 2013.

\bibitem{arraystore}
E.~Soroush, M.~Balazinska, and D.~L. Wang.
\newblock {ArrayStore: A Storage Manager for Complex Parallel Array
  Processing}.
\newblock In {\em {Proceedings of 2011 ACM SIGMOD International Conference on
  Management of Data}}, pages 253--264, 2011.

\bibitem{scidb:ssdbm-11}
M.~Stonebraker, P.~Brown, A.~Poliakov, and S.~Raman.
\newblock {The Architecture of SciDB}.
\newblock In {\em Proceedings of 2011 SSDBM International Conference on
  Scientific and Statistical Database Management}, pages 1--16, 2011.

\bibitem{sloan-sky-survey}
A.~S. Szalay and al.
\newblock {Designing and Mining Multi-Terabyte Astronomy Archives: the Sloan
  Digital Sky Survey}.
\newblock {\em SIGMOD Rec.}, 29(2), 2000.

\bibitem{ram-distributed}
A.~van Ballegooij, R.~Cornacchia, A.~P. de~Vries, and M.~Kersten.
\newblock {Distribution Rules for Array Database Queries}.
\newblock In {\em {Proceedings of 2005 DEXA International Conference on
  Database and Expert Systems Applications}}, pages 55--64, 2005.

\bibitem{ram}
A.~R. van Ballegooij.
\newblock {RAM: A Multidimensional Array DBMS}.
\newblock In {\em {Proceedings of 2004 EDBT Extended Database Technology
  Workshops}}, pages 154--165, 2004.

\bibitem{scimate}
Y.~Wang, W.~Jiang, and G.~Agrawal.
\newblock {SciMATE: A Novel MapReduce-Like Framework for Multiple Scientific
  Data Formats}.
\newblock In {\em {Proceedings of 2012 IEEE/ACM CCGRID International Symposium
  on Cluster, Cloud and Grid Computing}}, pages 443--450, 2012.

\bibitem{rasdaman:execution}
N.~Widmann and P.~Baumann.
\newblock {Efficient Execution of Operations in a DBMS for Multidimensional
  Arrays}.
\newblock In {\em Proceedings of 1998 SSDBM International Conference on
  Scientific and Statistical Database Management}, pages 155--165, 1998.

\bibitem{subzero}
E.~Wu, S.~Madden, and M.~Stonebraker.
\newblock {SubZero: A Fine-Grained Lineage System for Scientific Databases}.
\newblock In {\em {Proceedings of 2013 IEEE ICDE International Conference on
  Data Engineering}}, 2013.

\bibitem{SciQL}
Y.~Zhang, M.~Kersten, M.~Ivanova, and N.~Nes.
\newblock {SciQL: Bridging the Gap between Science and Relational DBMS}.
\newblock In {\em {Proceedings of 2011 IDEAS Symposium on International
  Database Engineering and Applications}}, pages 124--133, 2011.

\end{thebibliography}

\end{document}